\documentclass[a4paper,12pt,twoside,reqno,nonamelimits]{article}
\usepackage{a4wide}
\usepackage{authblk}
\usepackage{amsmath,amssymb,amsfonts,newlfont,amscd,amsthm,amsxtra,mathrsfs,nicefrac}
\usepackage[mathscr]{euscript}

\usepackage[utf8]{inputenc}
\usepackage{lmodern}
\usepackage[pagebackref,colorlinks=true,urlcolor=blue,linkcolor=blue,citecolor=blue]{hyperref}
\usepackage{url}
\usepackage{booktabs}
\usepackage{float,subfigure,graphicx}

\theoremstyle{definition}
\newtheorem{assum}{Assumption}[section]

\newcommand{\R}[1]{\widetilde{#1}}
\newcommand{\RV}[1]{\mathbf #1}
\newcommand{\Min}[1]{\min\limits_}
\newcommand{\Max}[1]{\max\limits_}
\DeclareMathOperator{\MI}{MI}
\DeclareMathOperator{\UI}{UI}
\DeclareMathOperator{\CI}{CI}
\DeclareMathOperator{\CoI}{CoI}
\DeclareMathOperator{\SI}{SI}

\title{\textsc{MaxEnt3D\_Pid}: An Estimator for the Maximum-entropy Trivariate Partial Information Decomposition}
\author[1]{Abdullah Makkeh}
\author[2,3]{Daniel Chicharro}
\author[1]{Dirk Oliver Theis}
\author[1]{Raul Vicente}
\affil[1]{\small Institute of Computer Science {\tiny of the } University of Tartu, Tartu, Estonia
}
\affil[2]{Department of Neurobiology, Harvard Medical School, Boston, MA, USA}
\affil[3]{Center for Neuroscience and Cognitive Systems @ UniTn, Istituto Italiano di Tecnologia, Rovereto (TN), Italy}
\begin{document}
    \maketitle
    \begin{abstract}
        Chicharro~\cite{Chicharro17b} introduced a procedure to determine multivariate partial information measures within the maximum entropy framework, separating unique, redundant, and synergistic components of information. Makkeh, Theis, and Vicente~\cite{makkeh2018broja} formulated the trivariate partial information measure of \cite{Chicharro17b} as Cone Programming. In this paper, we present \textsc{MaxEnt3D\_Pid}, a production-quality software that computes the trivariate partial information measure based on the Cone Programming model. We describe in detail our software, explain how to use it, and perform some experiments reflecting its accuracy in estimating the trivariate partial information decomposition.\newline  
    \end{abstract}
    
    \section{Introduction: Motivation and Significance} \label{sec:intro}
        The characterization of dependencies within complex multivariate systems helps identifying the mechanisms operating in the system and understanding their function. Recent work has developed methods to characterize multivariate interactions by separating $n$-variate dependencies for different orders $n$ \cite{Amari01, Schneidman03b, Timme14, Olbrich15, Perrone16}. In particular, the work of \cite{Williams10, Williams10b} introduced a framework, called Partial Information Decomposition (PID), which quantifies whether different input variables provide redundant, unique, or synergistic information about an output variable when combined with other input variables. Intuitively, inputs are redundant if each carries individually information about the same aspects of the output. Information is unique if it is not carried by any other single (or group of) variables, and synergistic information can only be retrieved combining several inputs.
        
        This information-theoretic approach to study interactions has found many applications to complex systems such as genes networks e.g.~\cite{Anastassiou07, Watkinson09, Chatterjee16}, interactive agents e.g.~\cite{Katz11, Flack2012, Ay2012, Frey18}, or neural processing e.g.~\cite{Marre09, Faes16, Pica17b}. More generally, the nature of the information contained in the inputs determines the complexity of extracting it \cite{Latham05, Steeg17}, how robust it is to disruptions of the system \cite{Rauh14b}, or how inputs dimensionality can be reduced without information loss \cite{Tishby99, Banerjee18c}.
        
        Despite this great potential, the applicability of the PID framework has been hindered by the lack
        of agreement on the definition of a suited measure of redundancy. In particular, \cite{Harder12} indicated that the original measure proposed by \cite{Williams10} only quantifies common amounts of information, instead of shared information that is qualitatively the same. A constellation of measures has been proposed to implement the PID e.g.~\cite{Harder12, Bertschinger12, Griffith13, Ince16, James17, Chicharro17, Finn18} and core
        properties, such as requiring nonnegativity as a property of the measures, are still a subject of debate \cite{Rauh17b, Finn18, James18c, Chicharro17c}
        
        A widespread application of the PID framework has also been limited by the lack of multivariate implementations. Some of the proposed measures were only defined for the bivariate case \cite{Harder12, Bertschinger12, Rauh17}. Other multivariate measures allow negative components in the PID \cite{Ince16, Finn18}, which, although it may be adequate for a statistical characterization of dependencies, limits the interpretation of the information-theoretic quantities in terms of information communication \cite{Cover06}. Among the PID measures proposed, the maximum entropy measures of \cite{Bertschinger12} have a preeminent role in the bivariate case because they provide bounds for any other measure consistent with a set of properties shared by many of the proposed measures. Motivated by this special role of the maximum entropy measures, \cite{Chicharro17b} extended the maximum entropy approach to measures of the multivariate redundant information, which provide analogous bounds for the multivariate case. However, \cite{Chicharro17b} did not address their numerical implementation.
        
        In this work we present \textsc{MaxEnt3D\_Pid}, a python module that computes a trivariate information decomposition following the maximum entropy PID of~\cite{Chicharro17b} and exploiting the connection with the bivariate decompositions associated with the trivariate one \cite{Chicharro17}. This is, to our knowledge, the first available implementation of the maximum-entropy PID framework beyond the bivariate case \cite{Makkeh17, Banerjee17, makkeh2018broja, dit}. This implementation is relevant for the theoretical development and practical use of the PID framework.
        
        From a theoretical point of view, this implementation will provide the possibility to test the properties of the PID beyond the bivariate case. This is critical with regard to the nonnegativity property because, while nonnegativity is guaranteed in the bivariate case, for the multivariate case it has been proven that negative terms can appear in the presence of deterministic dependencies \cite{Bertschinger12b, Rauh17b, Chicharro17c}. However, the violation of nonnegativity has only been proven with isolated counterexamples and it is not understood which properties of a system's dependencies lead to negative PID measures.
        
        From a practical point of view, the trivariate PID allows studying new types of distributed information that only appear beyond the bivariate case, such as information that is redundant to two inputs and unique with respect to a third \cite{Williams10}. This extension is significant both to directly study multivariate systems, as well as to be exploited for data analysis \cite{Tishby99, Burnham2002}. As mentioned above, the characterization of synergy and redundancy in multivariate systems is relevant for a broad range of fields that encompass social and biological systems. So far the PID  has particularly found applications in neuroscience e.g.~\cite{Stramaglia16, Wibral17, Ghazi17, Pica17b, Pica17, Faes17}. For data analysis, the quantification of multivariate redundancy can be applied to dimensionality reduction \cite{Banerjee18c} or to better understand how representations emerge in neural networks during learning \cite{Tax17, Schwartz17}. Altogether, this software promises to significantly contribute to the refinement of the information-theoretic tools it implements and also to foster its widespread application to analyze data from multivariate systems.
    
    \section{Models and software}\label{sec:soft-desc}
        The section starts by briefly describing the mathematical model of the problem. Then it discusses the architecture of the \textsc{MaxEnt3D\_Pid}. It closes by explaining in details how to use the software. 
        
        \subsection{Maximum Entropy Decomposition Measure}\label{subsec:the-measure}
            Consider $\RV{X},\RV{Y},$ and $\RV{Z}$ as the sources and $\RV{T}$ as the target of some system. Let $P$ be the joint distribution of $(\RV{T},\RV{X},\RV{Y},\RV{Z})$ and $\MI(\RV{T};\mathcal S)$ be the mutual information of $\RV{T}$ and $\mathcal S,$ where $\mathcal S$ is any nonempty subset of $(\RV{X},\RV{Y},\RV{Z})$. The PID decomposes $\MI(\RV{T};\RV{X},\RV{Y},\RV{Z})$ into finer parts, namely, synergistic, unique, redundant unique, and redundant information  These finer parts respect certain identities~\cite{Williams10}, e.g., a subset of them sums up to $\MI(\RV{T},\RV{X})$ (All identities are explained in Appendices~\ref{sec:apx-pid-frame} and~\ref{sec:apx-tri-pid}.). Following the maximum entropy approach \cite{Bertschinger12}, to obtain this decomposition, it is needed to solve the following optimization problems
            \begin{subequations}\label{eq:opt-pid}
                \begin{align}
                \Min_{\Delta_P}\MI(\RV{T};\RV{X},\RV{Y},\RV{Z})&~\label{eq:opt-I}\\
                \Min_{ \Delta_P}\MI(\RV{T};\RV{X_1},\RV{X_2})&\quad\text{for}~\RV{X_1},\RV{X_2}\in\{\RV{X},\RV{Y},\RV{Z}\}\label{eq:opt-II}
                \end{align}
            \end{subequations}
            where
            \begin{align*}
                \Delta_P= \{ Q\in\Delta: &Q(\RV{T},\RV{X}) = P(\RV{T},\RV{X}), Q(\RV{T},\RV{Y}) = P(\RV{T},\RV{Y}),\\ &Q(\RV{T},\RV{Z}) = P(\RV{T},\RV{Z})
                            \}
            \end{align*}
            and $\Delta$ is the set of all joint distributions of $(\RV{T},\RV{X},\RV{Y},\RV{Z})$. The four minimization problems in~\eqref{eq:opt-pid} can be formulated as exponential cone programs, a special of convex optimization. The authors refer to~\cite{makkeh2018broja} for a nutshell introduction to Cone programs, in particular, the exponential ones. The full details on how to formulate~\eqref{eq:opt-pid} as exponential cone programs and their convergence properties are explained in~\cite[Chapter 5]{makkeh2018applications}. 
            
            The \textsc{MaxEnt3D\_Pid} on its own returns the synergistic information and unique information collectively. In addition, with the help of the bivariate solver~\cite{Makkeh17} (used in a specific way) the finer synergistic and unique information can also be extracted. Hence, the presented model obtains all the trivariate PID quantities. The full details for recovering the finer parts can be found in Appendices~\ref{sec:apx-tri-pid} and~\ref{sec:apx-tri-fine}.
            
            \subsection{Software Architecture and Functionality}\label{subsec:soft-arch}
            
            \textsc{MaxEnt3D\_Pid} is implemented using the standard \textsc{Python} syntax. The module uses an optimization software \textsc{ECOS}~\cite{bib:Domahidi2013ecos} to solve several optimization problems needed to compute the trivariate PID. To install the module, \textsc{ECOS} python package has to be installed~\cite{ecos:github:18} and then from the \textsc{GitHub} repository the files \verb|MAXENT3D_PID.py|, \verb|TRIVARIATE_SYN.py|, \verb|TRIVARIATE_UNQ.py|, and \verb|TRIVARIATE_QP.py| must be downloaded~\cite{makkeh:github:chicharropid2018}. 
            
            \textsc{MaxEnt3D\_Pid} has two python classes \verb|Solve_w_ECOS| and \verb|QP|. Class \verb|Solve_w_ECOS| receives 
            the marginal distributions of $(\RV{T},\RV{X})$, $(\RV{T},\RV{Y})$, and $(\RV{T},\RV{Z})$ as python dictionaries. These distributions are used by \verb|Solve_w_ECOS| sub-classes \verb|Opt_I| and \verb|Opt_II| to solve the optimization problems of~\eqref{eq:opt-I} and~\eqref{eq:opt-II} respectively. The class \verb|QP| is used to recover the solution of any optimization problems of~\eqref{eq:opt-pid} when \verb|Solve_w_ECOS| fails to obtain a solution with a good quality. Figure~\ref{fig:flow_chart} gives an overview of how these two classes interact.
        
        \subsubsection[The Subclass Opt-I and Opt-II]{The Subclass \texttt{Opt\_I} and \texttt{Opt\_II}}
\begin{figure}
            \centering
            \rotatebox[origin=c]{270}{\includegraphics[width=20cm]{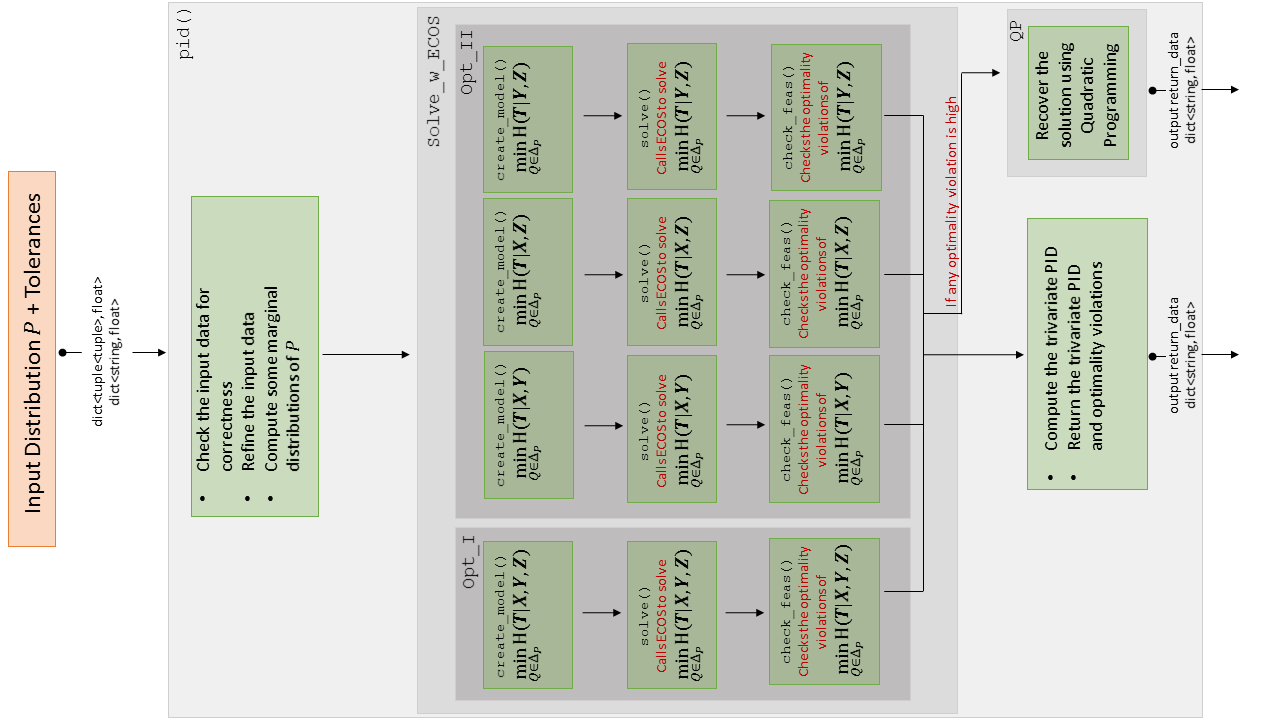}}
            \caption{\label{fig:flow_chart} A flow chart describing the process of computing the trivariate PID via \textsc{MaxEnt3D\_Pid}. It gives an overview of how \texttt{pid()} utilizes the classes \texttt{Solve\_w\_ECOS} and \texttt{QP} in the aim of computing the trivariate PID.}
        \end{figure}

            The sub-classes \verb|Opt_I| and \verb|Opt_II| formulate the problems~\eqref{eq:opt-pid}, use \textsc{ECOS} to get the optimal values, and compute their violations of the optimality certificates. They return the optimal values and their optimality violations. These violations are quality measures of the obtained PID. Figure~\ref{fig:flow_chart} describes this process within the class \verb|Solve_w_ECOS|. Note that both sub-classes \verb|Opt_I| and \verb|Opt_II| optimize conditional entropy functionals, however, the different number of arguments leads to a difference in how to fit the problems into the cone program and retrieving the optimal solution. Hence, the requirement of splitting them into different classes.
        
        \subsubsection[The Class QP]{The Class \texttt{QP}}
            Class \verb|QP| acts if \verb|Solve_w_ECOS| returns a values of a subset of~\eqref{eq:opt-pid} with high optimality violations. It improves the errand values by best fitting them using Quadratic Programming where the PID identities~\eqref{eq:ids-multi} are respected.
        \subsection[Using MAXENT3D-PID]{Using \textsc{MaxEnt3D\_Pid}}
            The process of computing the PID is packed in the function \verb|pid()|. This function takes as input the distribution $P$ of $(\RV{T},\RV{X},\RV{Y},\RV{Z})$ via a python dictionary where the tuples $(t,x,y,z)$ are keys and their associated probability $P(t,x,y,z)$ is the value of the key, see Figure~\ref{fig:code-pdf}. The function formulates and solves the problems of~\eqref{eq:opt-pid} using~\verb|Solve_w_ECOS|, and
            if needed uses \verb|QP| to improve the solution. This function~\verb|pid()| returns a python dictionary, explained in Table~\ref{tab:res} and Table~\ref{tab:vio}, containing the PID of $(\RV{T},\RV{X},\RV{Y},\RV{Z})$ in addition to the optimality violations.
            \begin{figure}[t!]
                \centering
                \begin{verbatim}
# The function pid() is imported from the module MAXENT3D_PID
from MAXENT3D_PID import pid

# The input distribution is defined as a python dictionary
andDgate = dict()
andDgate[ (0,0,0,0) ] = .25
andDgate[ (0,0,1,0) ] = .25
andDgate[ (0,1,0,1) ] = .25
andDgate[ (1,1,1,1) ] = .25

# pid() is called 
sol = pid(andDgate)
# printing the obtained PID
msg="""Synergistic information: {CI}
Unique information in X: {UIX}
Unique information in Y: {UIY}
Unique information in Z: {UIZ}
Unique information in X,Y: {UIXY}
Unique information in X,Z: {UIXZ}
Unique information in Y,Z: {UIYZ}
Shared information: {SI}"""
print(msg.format(**sol))
                \end{verbatim}
                \caption{\label{fig:code-pdf} Using~\textsc{MaxEnt3D\_Pid} to compute the PID of the distribution obtained from the~\textsc{AndDuplicate} gate (\texttt{andDgate}). \textsc{AndDuplicate} gate evalutes $\RV{T}$ as the logical and of $\RV{X}$ and $\RV{Y}$ ($\RV{X}\land\RV{Y}$) such that $\RV{Z}$ copies $\RV{X}$.
                }
            \end{figure}
            \begin{table}[t!]
                \centering
                \begin{tabular}{cc cc cc}
                    \multicolumn{2}{c}{{{ }}} 
                    & \multicolumn{2}{c}{{{ }}}\\
                    \toprule
                    Keys & Values & Keys & Values\\
                    \midrule
                    \verb|'UIX'|   & $\UI(\RV{T};\RV{X}\backslash\RV{Y},\RV{Z})$ &
                    \verb|'UIYZ'|  & $\UI(\RV{T};\RV{Y},\RV{Z}\backslash\RV{X})$\\
                    \verb|'UIY'|   & $\UI(\RV{T};\RV{Y}\backslash\RV{X},\RV{Z})$ &
                    \verb|'UIXZ'|  & $\UI(\RV{T};\RV{X},\RV{Z}\backslash\RV{Y})$\\
                     
                    \verb|'UIZ'|   & $\UI(\RV{T};\RV{Z}\backslash\RV{X},\RV{Y})$ &
                    \verb|'UIXY'|  & $\UI(\RV{T};\RV{X},\RV{Y}\backslash\RV{Z})$\\
                              
                    \verb|'CI'|    & $\CI(\RV{T};\RV{X},\RV{Y},\RV{Z})$ &
                    \verb|'SI'|    & $\SI(\RV{T};\RV{X},\RV{Y},\RV{Z})$\\
                    \bottomrule          
                \end{tabular}
                \caption{\label{tab:res} The keys of the trivariate PID quantities in the returned dictionary. Note that $\UI(\RV{T};\RV{X}_i\backslash\RV{X}_j,\RV{X}_k)$ and $\UI(\RV{T};\RV{X}_i,\RV{X}_k\backslash\RV{X}_j)$ refer to unique and unique redundant information for $\RV{X}_i,\RV{X}_k,\RV{X}_j\in\{\RV{X},\RV{Y},\RV{Z}\}$, $\CI(\RV{T};\RV{X},\RV{Y},\RV{Z})$ refers to synergistic information, and $\SI(\RV{T};\RV{X},\RV{Y},\RV{Z})$ refers to redundant or shared information.}
            \end{table}
            \begin{table}[t!]
                \centering
                \begin{tabular}{l l}
                    \toprule
                    Key & Value \\
                    \midrule
                    \verb|'Num_Err_I'|     & Optimality violations of $\Min_{ \Delta_P}\MI(\RV{T};\RV{X},\RV{Y},\RV{Z})$\\         
                    \verb|'Num_Err_12'|    & Optimality violations of $\Min_{ \Delta_P}\MI(\RV{T};\RV{X},\RV{Y})$\\
                    \verb|'Num_Err_13'|    & Optimality violations of $\Min_{ \Delta_P}\MI(\RV{T};\RV{X},\RV{Z})$\\
                    \verb|'Num_Err_23'|    & Optimality violations of $\Min_{ \Delta_P}\MI(\RV{T};\RV{Y},\RV{Z})$\\
                    \bottomrule
                \end{tabular}
                \caption{\label{tab:vio} The keys of optimality violations for each problem~\eqref{eq:opt-pid} in the returned dictionary.}
            \end{table}

            The function~\verb|pid()| has three other optional inputs. The first optional input is called \verb|parallel| (default value is~\verb|parallel='off'|) which determines whether the process will be parallelized. If \verb|parallel='off'|, then the process is going to be done sequentially, i.e., the four problems of~\eqref{eq:opt-pid} are going to be formulated and solved one after the other. Their optimality violations are also computed consecutively, and then final results are obtained. Whereas, when \verb|parallel='on'|, the formulation of the four problems~\eqref{eq:opt-pid} is done in parallel. The four problems are solved simultaneously, and finally the optimality violations along with the final results are computed in parallel. Thus, when \verb|parallel='on'| there will be three sequential steps: formulating the problems, solving them, and obtaining the final results as opposed to \verb|parallel='off'| which requires at least twelve sequential steps.
            
            The second optional input is a dictionary which allows the user to tune the tolerances controlling the optimization routines of~\textsc{ECOS} listed in Table~\ref{tab:ecos-parms}. 
            \begin{table}[t!]
            	\centering
            	\scalebox{0.9}[0.9]{
            	\begin{tabular}{llc}
            	\toprule
            	Parameter 				& Description 												& Default Value \\
            	\midrule
            	\texttt{feastol}		& 	primal/dual feasibility tolerance  						& $10^{-7}$         \\
            	\texttt{abstol}			&	absolute tolerance on duality gap						& $10^{-6}$         \\
            	\texttt{reltol}			&	relative tolerance on duality gap						& $10^{-6}$         \\
            	\texttt{feastol\_inacc} &	primal/dual infeasibility \textit{relaxed} tolerance	& $10^{-3}$         \\
            	\texttt{abstol\_inacc}	&	absolute \textit{relaxed} tolerance on duality gap		& $10^{-4}$         \\
            	\texttt{reltol\_inacc}	&	\textit{relaxed} relative duality gap					& $10^{-4}$         \\	
            	\texttt{max\_iter}		&	maximum number of iterations that \textsc{ECOS} does			& $100$             \\ 
            	\bottomrule
            	\end{tabular}%
            	}%
            	\caption{\label{tab:ecos-parms}Parameters (tolerances) that governs the optimization in \textsc{ECOS}.}
            \end{table}
            In this dictionary, the user only sets the parameters that will be tuned. For example, if the user wants to achieve high accuracy, then the parameters~\texttt{abstol} and \texttt{reltol} should be small (e.g.~$10^{-12}$) and the parameter~\texttt{max\_iter} should be high (e.g.~1000). In Figure~\ref{fig:code-ecos}, it is shown how to modify the parameters. In this case the solver will take longer to return the solution. For further details about parameter's tuning, check~\cite{makkeh2018broja}.
            \begin{figure}[t!]
                \centering
                \begin{verbatim}
# The function pid() is imported from the module MAXENT3D_PID

from MAXENT3D_PID import pid

# The input distribution is defined as a python dictionary
andDgate = dict()
andDgate[ (0,0,0,0) ] = .25
andDgate[ (0,0,1,0) ] = .25
andDgate[ (0,1,0,1) ] = .25
andDgate[ (1,1,1,1) ] = .25

# The dictionary is defined to tune ECOS parameters
parms = dict()
# abstol is set
parms['abstol'] = 1.e-12
# reltol is set 
parms['reltol'] = 1.e-12
# max_iter is set 
parms['max_iters'] = 100
# pid is called
pid(andDgate, parallel='on', **parms)
                    \end{verbatim}
                \caption{\label{fig:code-ecos} Tuning the parameters of \textsc{ECOS}}
            \end{figure}
            The third optional input is called \verb|output| and it controls what will~\verb|pid()| print on the user's screen. This optional input is explained in Table~\ref{tab:printing-mode}.  
            \begin{table}[t!]
            	\centering
            	\scalebox{0.9}[0.9]{
            	\begin{tabular}{ll}
            		\toprule
            		Value	& Description 																	\\
            		\midrule
            		0 (default)		& {\bf Simple Mode:} \texttt{pid()} prints its output (python dictionary).\\
            		1				& {\bf Time Mode:} In addition to what is printed when \texttt{output=0},\\
            		                &\texttt{pid()} prints a flag when it starts preparing the optimization problems\\
            		                &in~\eqref{eq:opt-pid}, the total time to create each problem, a flag when it calls~\textsc{ECOS},\\
            	     				& brief stats from~\textsc{ECOS} of each problem after solving it (Figure~\ref{fig:code-brief}),	\\
            	     				& the total time for retrieving the results, the total time for computing\\
            	     				&the optimality violations, and the total time to store the results.\\
            	     				&\\
            		2				& {\bf Detailed Time Mode:} In addition to what is printed when \texttt{output=0},\\ 
            		                &\texttt{pid()} prints for each problem the time of each major step of creating\\
            		                &the model, brief stats from~\textsc{ECOS} of each problem after solving it,	\\
            		                & the total time of each function used for retrieving the results,\\
            		                & the time of each major step used to computing the optimality violations,\\
            		                & the time of each function used to obtain the final results,\\
            		                & and the total time to store the results.\\
            		                &\\
            		3				& {\bf Detailed Optimization Mode:} In addition to what is printed when\\
            		                &\texttt{output=1}, \texttt{pid()} prints \textsc{ECOS} detailed stats of each problem\\
            		                &after solving it (Figure~\ref{fig:ecos-full}).\\
            		\bottomrule
            	\end{tabular}%
            	}%
            	\caption{\label{tab:printing-mode} Description of the printing modes in the function \texttt{pid()}.}
            \end{table}
            \begin{figure}[t!]
                \centering
                \footnotesize
                \begin{verbatim}
MAXENT3D_PID.pid(): Stats for optimizing H(S|X,Y,Z):
{'exitFlag': 0, 'pcost': -0.3465735936653011, 'dcost': -0.3465735930504127,
'pres': 2.8092696654527348e-09, 'dres': 2.7131747078266765e-10, 'pinf': 0.0,
'dinf': 0.0, 'pinfres': nan, 'dinfres': 0.43220650722288695, 'gap':
6.4243546975262245e-09, 'relgap': 1.8536769144998566e-08, 'r0': 1e-08, 
'iter': 19, 'mi_iter': -1, 'infostring': 'Optimal solution found', 'timing':
{'runtime': 0.000498888, 'tsetup': 9.6448e-05, 'tsolve': 0.00040244}, 
'numerr': 0}
                \end{verbatim}
                \caption{\label{fig:code-brief} Brief stats from~\textsc{ECOS} after solving problem~\eqref{eq:opt-I}}
            \end{figure}
            
            \begin{figure}[t!]
                \footnotesize
                \begin{verbatim}
ECOS 2.0.4 - (C) embotech GmbH, Zurich Switzerland, 2012-15. Web: www.embotech.com/ECOS

It     pcost       dcost      gap   pres   dres    k/t    mu     step   sigma     IR    |   BT
 0  +0.000e+00  -0.000e+00  +3e+01  1e+00  3e-01  1e+00  1e+00    ---    ---    0  0  - |  -  -
 1  -1.622e+00  -1.086e+00  +6e+00  7e-01  1e-01  1e+00  2e-01  0.7833  9e-03   1  1  1 |  1  1
 2  -1.546e+00  -1.369e+00  +1e+00  3e-01  3e-02  3e-01  5e-02  0.7833  9e-03   1  1  1 |  1  1
 3  -7.067e-01  -6.684e-01  +3e-01  9e-02  7e-03  6e-02  1e-02  0.7833  1e-02   1  1  1 |  1  1
 4  -4.724e-01  -4.501e-01  +1e-01  5e-02  5e-03  4e-02  6e-03  0.5013  2e-01   1  1  1 |  4  3
 5  -3.842e-01  -3.755e-01  +6e-02  2e-02  2e-03  1e-02  2e-03  0.6266  5e-02   1  1  1 |  2  2
 6  -3.588e-01  -3.567e-01  +1e-02  5e-03  5e-04  4e-03  6e-04  0.7833  5e-02   1  1  1 |  2  1
 7  -3.509e-01  -3.499e-01  +7e-03  3e-03  2e-04  2e-03  3e-04  0.6266  2e-01   1  1  1 |  4  2
 8  -3.481e-01  -3.479e-01  +2e-03  9e-04  8e-05  5e-04  8e-05  0.9791  3e-01   1  1  1 |  5  0
 9  -3.468e-01  -3.467e-01  +6e-04  2e-04  2e-05  1e-04  2e-05  0.7833  5e-02   1  1  1 |  2  1
10  -3.466e-01  -3.466e-01  +2e-04  9e-05  9e-06  5e-05  8e-06  0.6266  5e-02   2  1  1 |  2  2
11  -3.466e-01  -3.466e-01  +5e-05  2e-05  2e-06  1e-05  2e-06  0.7833  1e-02   1  1  1 |  1  1
12  -3.466e-01  -3.466e-01  +2e-05  8e-06  8e-07  4e-06  7e-07  0.7833  2e-01   1  0  0 |  4  1
13  -3.466e-01  -3.466e-01  +8e-06  3e-06  3e-07  2e-06  3e-07  0.6266  5e-02   2  0  0 |  2  2
14  -3.466e-01  -3.466e-01  +2e-06  9e-07  8e-08  4e-07  7e-08  0.7833  5e-02   2  0  0 |  2  1
15  -3.466e-01  -3.466e-01  +8e-07  3e-07  3e-08  2e-07  3e-08  0.6266  5e-02   1  0  0 |  2  2
16  -3.466e-01  -3.466e-01  +2e-07  8e-08  7e-09  4e-08  7e-09  0.7833  9e-03   2  0  0 |  1  1
17  -3.466e-01  -3.466e-01  +7e-08  3e-08  3e-09  2e-08  3e-09  0.6266  5e-02   2  0  0 |  2  2
18  -3.466e-01  -3.466e-01  +2e-08  7e-09  7e-10  4e-09  6e-10  0.7833  9e-03   1  0  0 |  1  1
19  -3.466e-01  -3.466e-01  +6e-09  3e-09  3e-10  1e-09  2e-10  0.6266  5e-02   0  0  0 |  2  2

OPTIMAL (within feastol=2.8e-09, reltol=1.9e-08, abstol=6.4e-09).
Runtime: 0.000499 seconds.
                \end{verbatim}
                \caption{\label{fig:ecos-full} Detailed stats from~\textsc{ECOS} after solving problem~\eqref{eq:opt-I}}
            \end{figure}

    \section{Illustrations} \label{sec:illustrations}
        This section shows some performance tests of \textsc{MaxEnt3D\_Pid} on three types of instances. We will describe each type of instances and show the results of testing \textsc{MaxEnt3D\_Pid} for each one of them. The first two types, paradigmatic and \textsc{Copy} gates, are used as validation and memory tests. The last type, random probability distributions, is used to evaluate the accuracy and efficiency of \textsc{MaxEnt3D\_Pid} in computing the trivariate partial information decomposition. The machine used comes with Intel(R) Core(TM) i7-4790K CPU (4 cores) and 16GB of RAM. Only the computations of the last type were done using parallelization. 
        
        \subsection{Paradigmatic Gates}\label{subsec:gates}
            As a first test, we use some trivariate PIDs that are known and have been studied previously~\cite{griffith2014quantifying}. These examples are the logic gates collected in Table~\ref{tab:gates}. For these examples the decomposition can be derived analytically and thus they serve to check the numerical estimations. 
            \begin{table}[t!]
                \centering\footnotesize
                \begin{tabular}{|l l|}
                    \toprule
                     Instance               & Operation \\
                     \midrule
                     \textsc{XorDuplicate}  & $\RV{T}=\RV{X}\oplus\RV{Y};\RV{Z} = \RV{X};\RV{X},\RV{Y}$ i.i.d.\\
                     \midrule
                     \textsc{XorLoses}      & $\RV{T}=\RV{X}\oplus\RV{Y};\RV{Z} = \RV{X}\oplus\RV{Y};\RV{X},\RV{Y}$ i.i.d.\\
                     \midrule
                     \textsc{XorMultiCoal}  & $\RV{T}=\RV{U}\oplus\RV{V}\oplus\RV{W}; \RV{X}=(\RV{U},\RV{V}),$\\
                                            & $\RV{Y}=(\RV{U},\RV{W}),\RV{Z}=(\RV{V},\RV{W}); \RV{U},\RV{V},\RV{W}$ i.i.d.\\
                    \midrule
                     \textsc{AndDuplicate}  & $\RV{T}=\RV{X}\land\RV{Y}; \RV{Z} = \RV{X};\RV{X},\RV{Y}$ i.i.d.\\
                     \bottomrule
                \end{tabular}
                \caption{\label{tab:gates} Paradigmatic gates with a brief explanation of their operation, where $\oplus$ is the logical~\textsc{Xor} and $\land$ is the logical~\textsc{And}.}
                
            \end{table}
        \subsubsection{Testing} 
            The test is implemented in \verb|test_gates.py|. \textsc{MaxEnt3D\_Pid} returns, for all gates, the same values as~\cite[][Table 1]{griffith2014quantifying} up to a precision error of order $10^{-9}$. The slowest solving time (not in parallel) is 1 millisecond.
        \subsection{Copy Gate}
            As a second test, we use the \textsc{Copy} gate example to examine the simulation of large systems. We show how the solver handles large systems in terms of speed and reliability. 
            
            The \textsc{Copy} gate is the mapping of $(x,y,z)$, chosen uniformly at random, to $(t,x, y, z)$ where
            $t = (x, y, z).$ The size of the joint distribution of $(\RV{T},\RV{X},\RV{Y},\RV{Z})$ scales as $|X|^2\cdot|Y|^2\cdot|Z|^2$ where $x,y, z\in X\times Y\times Z.$ In our test, $|X|=\ell,|Y|=m$ and $|Z|=n$ where $2\le \ell,m,n\le 50$.   
            
            Since $\RV{X},\RV{Y}$ and $\RV{Z}$ are independent, it is easy to see that the only nonzero quantities are $\UI(\RV{T};\RV{X_1}\backslash\RV{X_2},\RV{X_3}) = H(\RV{X_1})$ for $\RV{X_1},\RV{X_2},\RV{X_3}\in\{\RV{X},\RV{Y},\RV{Z}\}$.
        \subsubsection{Testing}
            The test is implemented in \verb|test_copy_gate.py|. The slowest solving time was 100 sec and the worst deviation from the actual values was $0.00001\%$.
        \subsection{Random Probability Distributions}\label{s3_3}
            As a last example we use joint distributions of $(\RV{T},\RV{X},\RV{Y}, \RV{Z})$ sampled uniformly at random over the probability space, to test the accuracy of the solver. The size of $T$, $X$, and $Y$ is fixed to 2 whereas $|Z|$ varies in $\{2,\dots,14\}$. For each $|Z|,$ 500 joint distributions of $(\RV{T},\RV{X},\RV{Y}, \RV{Z})$ are sampled.
        \subsubsection{Testing}
\begin{figure}[t!]
                \centering
                \subfigure[$\UI(\RV{T};\RV{X}\backslash\RV{Y},\RV{Z})$]{\includegraphics[width=7cm, height=6cm]{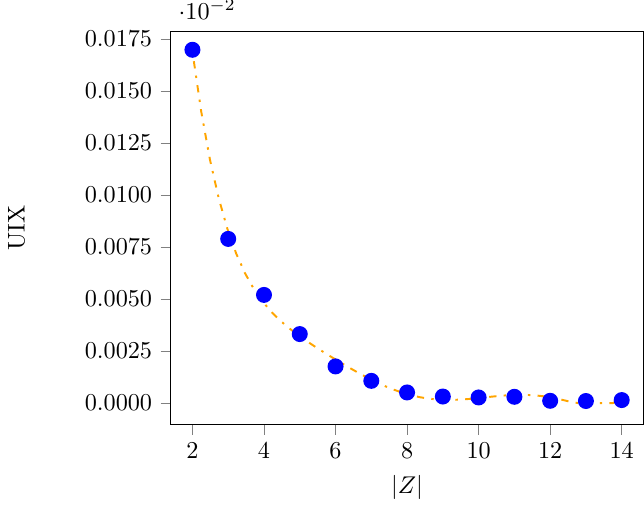}}
            	\subfigure[$\UI(\RV{T};\RV{Y}\backslash\RV{X},\RV{Z})$]{\includegraphics[width=7cm, height=6cm]{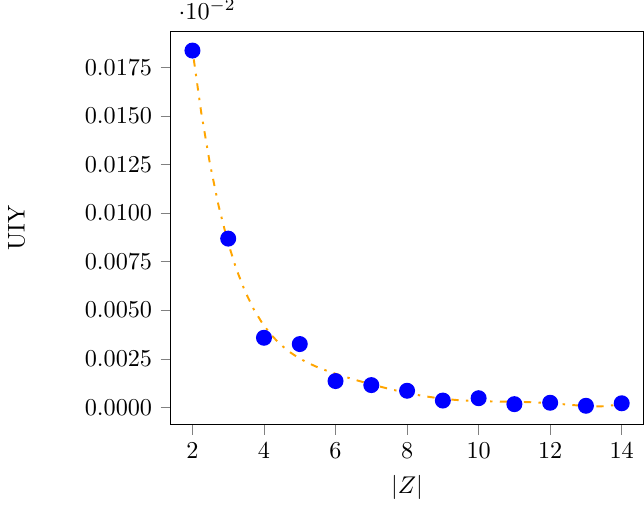}}
            	\subfigure[$\UI(\RV{T};\RV{Z}\backslash\RV{X},\RV{Y})$]{\includegraphics[width=7cm, height=6cm]{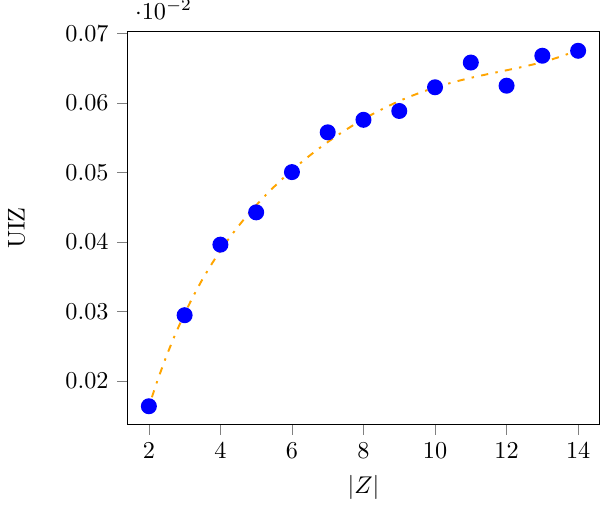}}
                \caption{The variation of the unique information, as the size of $Z$ increases, for the random probability distributions described in Section \ref{s3_3}. It shows that the value of unique information of $\RV{Z}$ increases as the dimension of $Z$ increases.}
                \label{fig:unique}
            \end{figure}

\begin{figure}[t!]
                \centering
                \includegraphics[width=10cm, height=8cm]{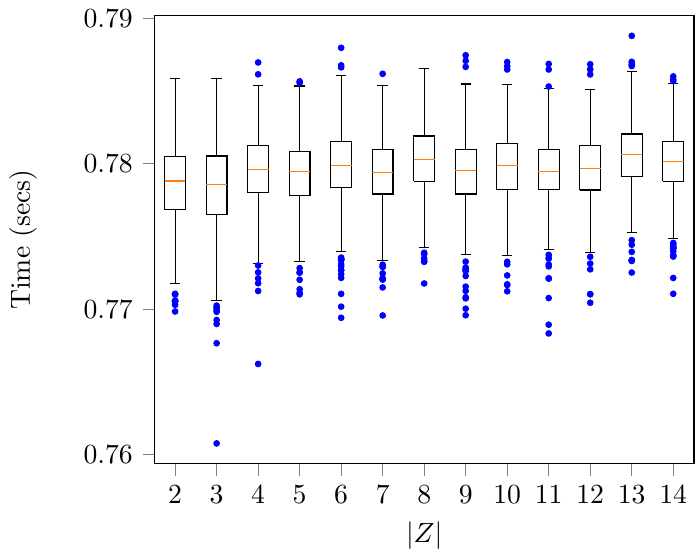}
                \caption{Box plotting of the time for \textsc{MaxEnt3D\_Pid} to compute the PID of a random joint probability distributions of $(\RV{T},\RV{X},\RV{Y}, \RV{Z})$ for $|T|=|X|=|Y|=2$ and different sizes of $Z$. For the size of the sets explored, the computational time shows a flat trend and its variance is small.}
                \label{fig:time}
            \end{figure}

            As $|Z|$ increases, the average value of $\UI(\RV{T};\RV{X}\backslash\RV{Y},\RV{Z})$ and of$\UI(\RV{T};\RV{Y}\backslash\RV{X},\RV{Z})$ decrease while that of $\UI(\RV{T};\RV{Z}\backslash\RV{X},\RV{Y})$ increases. In Figure~\ref{fig:unique}, the accuracy of the optimization is reflected in the low divergence from zero obtained for the unique information $\UI(\RV{T};\RV{X}\backslash\RV{Y},\RV{Z})$ and $\UI(\RV{T};\RV{Y}\backslash\RV{X},\RV{Z})$. In Figure~\ref{fig:time}, it the time has a constant trend and the highest time value recorded is $0.8$ sec.
    \section{Summary and discussion} \label{sec:summary}

        In this work we presented \textsc{MaxEnt3D\_Pid}, a python module that computes a trivariate decomposition based on the Partial Information Decomposition (PID) framework of \cite{Williams10}, in particular following the maximum entropy PID of~\cite{Chicharro17b} and exploiting the connection with the bivariate decompositions associated with the trivariate one \cite{Chicharro17}. This is, to our knowledge, the first available implementation extending the maximum-entropy PID framework beyond the bivariate case \cite{Makkeh17, Banerjee17, makkeh2018broja, dit}.
        
        The PID framework allows decomposing the information that a group of input variables has about a target variable into redundant, unique, and synergistic components. For the bivariate case, this results in a decomposition with four components, quantifying the redundancy, synergy, and unique information of each of the two inputs. In the multivariate case, finer parts appear which do not correspond to purely redundant or unique components. For example, the redundancy components of the multivariate decomposition can be interpreted based on local unfoldings when a new input is added, with each redundancy component unfolding into a component also redundant with the new variable and a component of unique redundancy with respect to it \cite{Chicharro17b}. The PID analysis can qualitatively characterize the distribution of information beyond the standard mutual information measures \cite{James16} and has already been proven useful to study information in multivariate systems e.g.~\cite{Lizier13, Wibral15, Banerjee15, James16, Pica17, Pica17b, Kay18, Frey18, Crosato17, Sootla17}.
        
        However, the definition of suited measures to quantify synergy and redundancy is still a subject of debate. From all proposed PID measures, the maximum entropy measures \cite{Bertschinger12} have a preeminent role in the bivariate case because they provide bounds to any other alternative measures that share fundamental properties related to the notions of redundancy and unique information. \cite{Chicharro17b} generalized the maximum entropy approach proposing multivariate definitions of redundant information and showing that these measures implement the local unfolding of redundancy via hierarchically related maximum entropy constraints. The package \textsc{MaxEnt3D\_Pid} efficiently implements the constrained information minimization operations involved in the calculation of the trivariate maximum-entropy PID decomposition. In Section~\ref{sec:soft-desc}, we described the architecture of the software, presented in details the main function of the software that computes the PID along with its optional inputs, and described how to use it. In Section \ref{sec:illustrations}, we provided examples which verified that the software produces correct results on paradigmatic gates, showed how the software scales with large systems, and reflected the accuracy of the software in estimating PID.
        
        The possibility to calculate a trivariate decomposition of the mutual information represents a qualitative extension of the PID framework that goes beyond an incremental extension of the bivariate case, both regarding its theoretical development and its applicability. From a theoretical point of view, regarding the maximum-entropy approach, the multivariate case requires the introduction of new types of constraints in the information minimization that do not appear in the bivariate case \cite[and Section~\ref{sec:soft-desc}]{Chicharro17b}. More generally, the trivariate decomposition allows further studying one of the key unsolved issues in the PID formulation, namely the requirement of nonnegativity of the PID measures in the multivariate case.
        
        In particular, \cite{Harder12} indicated that the original measure proposed by \cite{Williams10} only quantifies common amounts of information, and required new properties for the PID measures, to quantify qualitatively and not quantitatively how information is distributed. However, for the multivariate case these properties have been proven to be incompatible with guaranteeing nonnegativity, by using some counterexamples \cite{Bertschinger12b, Rauh17b, Chicharro17c}. This led some subsequent proposals to define PID measures that either focus on the bivariate case \cite{Harder12, Bertschinger12} or do not require nonnegativity \cite{Ince16, Finn18}. A multivariate formulation is desirable because the notions of synergy and redundancy are not restrained to the bivariate case, while nonnegativity is required for an interpretation of the measures in terms of information communication \cite{Cover06} and not only as a statistical description of the probability distributions. \textsc{MaxEnt3D\_Pid} will allow systematically exploring when negative terms appear, beyond the currently studied isolated counterexamples. Furthermore, it has been shown that in those counterexamples negative terms result from the criterion used to assign identity to different pieces of information when deterministic relations exist \cite{Chicharro17c}. Therefore, a systematic analysis of the appearance of negative terms will provide a better understanding of how information identity is assigned when quantifying redundancy, which is fundamental to assess how the PID measures conform to the corresponding underlying concepts.
        
        From a practical point of view, the trivariate decomposition allows studying qualitatively new types of distributed information, identifying finer parts of the information that the inputs have about the target, such as information that is redundant to two inputs and unique with respect to a third \cite{Williams10}. This is particularly useful when examining multivariate representations, such as the interactions between several genes \cite{Anastassiou07, Erwin09} or characterizing the nature of coding in neural populations \cite{Olshausen97, Palmer15}. Furthermore, exploiting the connection between the bivariates and the trivariate decomposition due to the invariance of redundancy to context \cite{Chicharro17}, \textsc{MaxEnt3D\_Pid} also allows estimating the finer parts of the synergy component (Appendix~\ref{sec:apx-tri-fine}). This also offers a substantial extension in the applicability of the PID framework, in particular for the study of dynamical systems \cite{Faes15, Chicharro12b}. In particular, a question that requires a trivariate decomposition is how information transfer is distributed among multivariate dynamic processes. Information transfer is commonly quantified with the measure called transfer entropy \cite{Schreiber00b, Vicente10, Hlavavckova07, Vicente14}, which calculates the conditional mutual information between the current state of a certain process $Y$ and the past of another process $X$, given the past of $Y$ and of any other processes $Z$ that may also influence those two. In this case, by construction, the PID analysis should operate with three inputs corresponding to the pasts of $X$, $Y$, and $Z$. Transfer entropy is widely applied to study information flows between brain areas to characterize dynamic functional connectivity \cite{Valdes11, Wibral14a, Wibral14c}, and characterizing the synergy, redundancy, and unique information of these flows can provide further information about the degree of integration or segregation across brain areas \cite{Deco15}. 
        
        More generally, the availability of a software implementing the maximum entropy PID framework beyond the bivariate case, promises to be useful in a wide range of fields in which interactions in multivariate systems are relevant, spanning the domain of social \cite{Flack2012, Daniels16} and biological sciences  \cite{Erwin09, Timme14, Chatterjee16, Pica17b}. Furthermore, the PID measures can also be used as a tool for data analysis and to characterize computational models. This comprises dimensionality reduction via synergy or redundancy minimization \cite{Steeg17, Banerjee18c}, the study of generative networks that emerge from information maximization constraints \cite{Linsker92, Bell95}, or explaining the representations in deep networks \cite{Schwartz17}.

        The \textsc{MaxEnt3D\_Pid} package presents several differences and advantages with respect to other software packages currently available to implement the PID framework. Regarding the maximum-entropy approach, other packages only compute bivariate decompositions \cite{Makkeh17, Banerjee17, makkeh2018broja, dit}. The dit package~\cite{dit} also implements several other PID measures, including bivariate implementations for the measure of \cite{James17} and \cite{Harder12}. Among the multivariate decompositions, the ones using the measures $I_{min}$~\cite{Williams10} or $I_{MMI}$~\cite{Barret15} can readily be calculated with standard estimators of the mutual information. However, the former, as discussed above, only quantifies common amounts of information, while the latter is only valid for a certain type of data, namely multivariate gaussian distributed. Software to estimate multivariate pointwise PIDs is also available \cite{Ince16, Finn18, Lizier14}~\footnote{ The Ince software is located at~\url{ https://github.com/robince/partial-info-decomp} and Finn and Lizier software at~\url{http://jlizier.github.io/jidt/}.}. However, as mentioned above, these measures by construction allow negative components, which may not be desirable for the interpretation of the decomposition and limits their applicability for data analysis \cite{Banerjee18c}. Altogether, \textsc{MaxEnt3D\_Pid} is the first software that implements the mutual information PID framework via hierarchically related maximum entropy constraints, extending the bivariate case by efficiently computing the trivariate PID measures.            

    \section*{Computational details}
        The results in this paper were obtained using \textsc{Python}~3.6.7 and the conic solver~\textsc{ECOS} 2.0.4. \textsc{Python} and all its packages are available at~\url{https://www.python.org/}.
    
    \section*{Acknowledgments}
        This research was supported by the Estonian Research Council, ETAG (Eesti Teadusagentuur), through PUT Exploratory Grant \#620. D.C. was supported by the Fondation Bertarelli. R.V. also thanks the financial support from ETAG through the personal research grant PUT1476. We also gratefully acknowledge funding by the European Regional Development Fund through the Estonian Center of Excellence in IT, EXCITE.


\newpage

\begin{appendix}
\section{Williams-Beer PID Framework}\label{sec:apx-pid-frame}	
    In order to decompose $\MI(\RV{T},\RV{S})$ where $\RV{T}$ is the target and $\RV{S}$ are the sources. \cite{Williams10} defined a set of axioms leading to what is known as the redundancy lattice (Figure~\ref{fig:lattice}). These axioms and lattice form the framework for partial information decomposition (PID) upon which all the exiting definitions of PID are formulated. 
    \subsection{Williams-Beer Axioms}
		Suppose that a \textit{source} $A$ is a subset of $\RV{S}$ and a \textit{collection} $\alpha$ is a set of sources. A shorthand notation inspired by~\cite{Chicharro17b} will be used to represent the collection of sources, for example, if the system is $(\RV{T},\RV{X},\RV{Y},\RV{Z}),$ then the collection of sources $\{\{X,Y\},\{X,Z\}\}$ will be denoted as $XY.XZ$. \cite{Williams10} defined the following axioms that redundancy should comply:
		\begin{itemize}
			\item Symmetry \textbf{(S)}: $\MI(\RV{T}; \alpha)$ is invariant to the order of the sources in the collection.
			\item Self-redundancy \textbf{(SR)}: The redundancy of a collection formed by a single source is equal to the mutual information of that source.
			\item  Monotonicity \textbf{(M)}: Adding sources to a collection can only decrease the redundancy of the resulting collection, and redundancy is kept constant when adding a superset of any of the existing sources.
		\end{itemize}
	\subsection{The Redundancy Lattice}
\begin{figure}[t!]
			\centering
			\includegraphics[height=8cm, width=16cm]{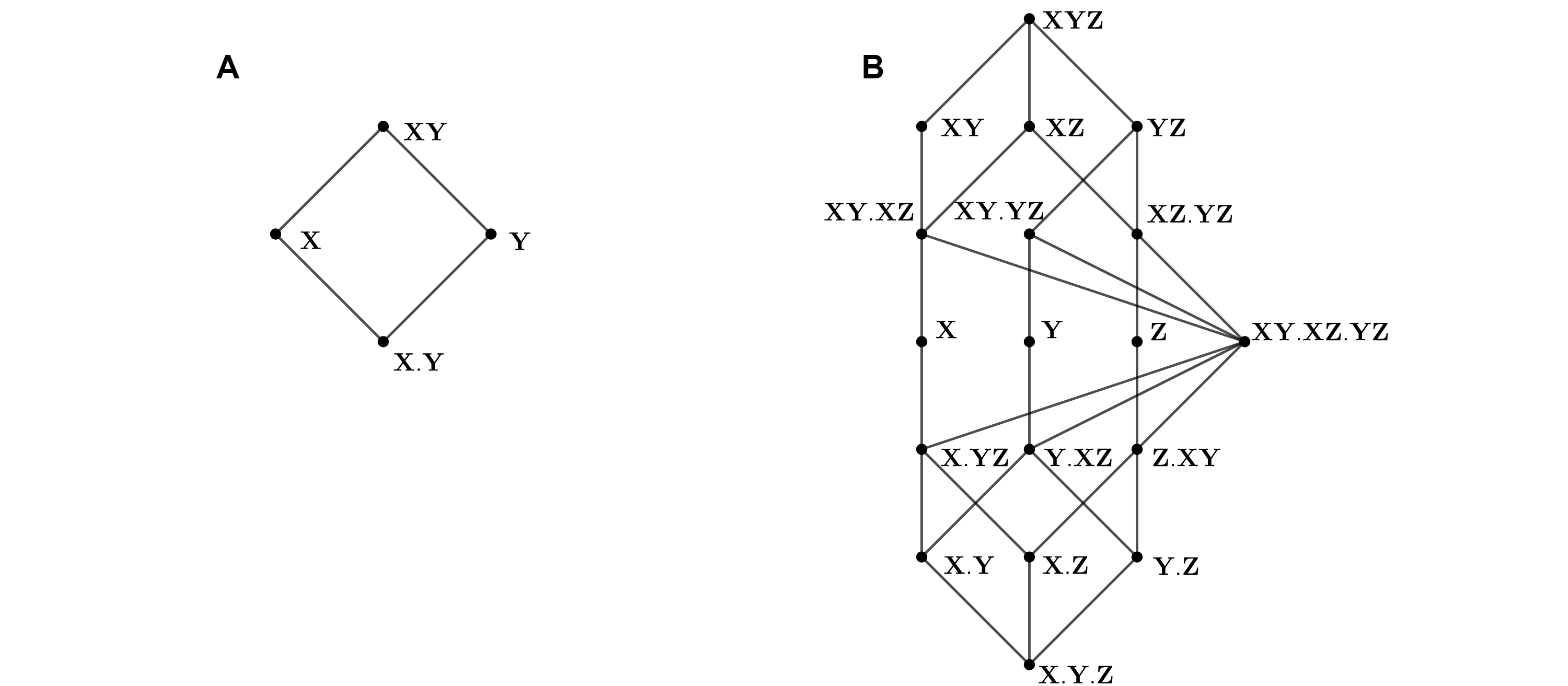}
			\caption{Bivariate and Trivariate redundancy Lattices. Letters indicate the mapping of terms between the lattices.}\label{fig:lattice}
		\end{figure}

	    \cite{Williams10} defined a lattice formed from the collections of sources. They used \textbf{(M)} to define the partial ordering between the collections. The axiom \textbf{(S)} reflects the fact that each atom of the lattice will represent a partial information decomposition quantity. More importantly, not all the collections of sources will be considered as atoms since adding a superset of any source to the examined system does not change redundancy, i.e.~, \textbf{(M)}. The set of collections of sources included in the lattice which will form its atoms is defined as:
		\begin{equation}
			\mathcal{A}(\RV{S}) = \{\alpha\in \mathcal{P}(\RV{S}) - \{\emptyset\}: \forall A_i,A_j\in\alpha, A_i\nsubseteq A_j\},
		\end{equation}
		where $\mathcal{P}(\RV{S})$ is the power set of $\RV{S}$. For this set of collections (atoms), the partial ordering relation that construct the Redundancy lattice is 
		\begin{equation}
		    \forall \alpha,\beta\in \mathcal{A}(\RV{S}),(\alpha\preceq\beta\Leftrightarrow\forall B\in\beta,\exists A\in\alpha,A\subseteq B),
		\end{equation}
		i.e.~, for two collections $\alpha$ and $\beta$, $\alpha\preceq\beta$ if for each source in $\beta$ there is a source in $\alpha$ that is a subset of that source. In Figure~\ref{fig:lattice}, the bivariate and trivariate redundancy lattices are shown. 
		\subsection{Defining PID over the Redundancy lattice}
		The mutual information decomposition was constructed in~\cite{Williams10} by implicitly defining partial information measures $\delta_\mathcal{C}(\RV{T};\alpha)$ associated with each node $\alpha$ of the redundancy lattice $\mathcal{C}$ (Figure~\ref{fig:lattice}), such that the redundancy measures are obtained as
		\begin{equation}
			\MI(\RV{T},\alpha) = \sum_{\beta\in\downarrow\alpha} \delta_\mathcal{C}(\RV{T};\beta),
		\end{equation}\label{eq:lat-main-def}
		where $\downarrow\alpha$ refers to the set of collections lower than or equal to $\alpha$ in the partial ordering, and hence reachable descending from $\alpha$ in the lattice $\mathcal{C}$.
\section{Bivariate Partial Information Decomposition}
	Let $\RV{T}$ be the target random variable, $\RV{X}$ and $\RV{Y}$ be the two source random variables, and $P$ be the joint probability distribution of $(\RV{T},\RV{X},\RV{Y})$. The PID captures the synergistic, unique, and redundant information as follows:
	\begin{itemize}
		\item The synergistic information between $\RV{X}$ and $\RV{Y}$ about $\RV{T}$, namely, $\CI(\RV{T};\RV{X}:\RV{Y})$. 
		\item The redundant information of $\RV{X}$ and $\RV{Y}$ about $\RV{T}$, namely, $\SI(\RV{T};\RV{X},\RV{Y})$.
		\item The unique information of $\RV{X}$ about $\RV{T}$, namely, $\UI(\RV{T};\RV{X}\backslash\RV{Y})$.
		\item The unique information of $\RV{Y}$ about $\RV{T}$, namely, $\UI(\RV{T};\RV{Y}\backslash\RV{X})$.
	\end{itemize}
	This decomposition, using Beer-Williams axioms, yields these identities:
	\begin{equation}
		\begin{split}
			\MI(\RV{T};\RV{X},\RV{Y}) &= \CI(\RV{T};\RV{X}:\RV{Y}) + \SI(\RV{T};\RV{X},\RV{Y}) + \UI(\RV{T};\RV{X}\backslash\RV{Y}) + \UI(\RV{T};\RV{X}\backslash\RV{Y})\\
			\MI(\RV{T};\RV{X}_i) &= \SI(\RV{T};\RV{X}_i,\RV{X}_j) + \UI(\RV{T};\RV{X}_i\backslash\RV{X}_j)\quad\text{for all}~\RV{X}_i,\RV{X}_j\in\{\RV{X},\RV{Y}\}.
		\end{split}
	\end{equation}
	Given the generic structure of the PID framework, \cite{Bertschinger12}~(BROJA) defined PID measures considering the following polytope:
		\begin{equation}
			\Delta_P = \{Q\in\Delta: Q(\RV{T},\RV{X}) = P(\RV{T},\RV{X}), Q(\RV{T},\RV{Y}) = P(\RV{T},\RV{Y})\},
		\end{equation}
	where $\Delta$ is the set of all joint distributions of $(\RV{T},\RV{X},\RV{Y})$. 
	 \cite{Bertschinger12}~(BROJA) used the maximum entropy decomposition over $\Delta_P$ in order to quantify the above quantities. Moreover, BROJA assumed that the following assumptions holds.
	\begin{assum}[Lemma 3~\cite{Bertschinger12}]\label{assum:bi}
		On the bivariate redundancy lattice (Figure~\ref{fig:lattice}), the following assumptions must hold to quantify the PID
		\begin{enumerate}
			\item All partial information measures of the redundancy lattice are nonnegative.\label{assm:bi-neg}
			\item The terms $\delta(\RV{T};\RV{X}.\RV{Y}), \delta(\RV{T};\RV{X}),$ and  $\delta(\RV{T};\RV{Y}),$ are constant on $\Delta_P$.\label{assm:bi-cst}
			\item The synergistic term, namely, $\delta(\RV{T},\RV{X}\RV{Y})$ vanishes on $\Delta_P$ upon minimizing the mutual information  $\MI(\RV{T};\RV{X},\RV{Y})$.\label{assm:bi-syn} 
		\end{enumerate}
	\end{assum}
	Under the above assumptions and using maximal entropy decomposition, BROJA defined the following optimization problems that computes the PID quantities.
	\begin{subequations}
		\begin{align}
			\R{\CI}(\RV{T};\RV{X}:\RV{Y})			&= \MI(\RV{T}; \RV{X},\RV{Y}) - \Min_{\tiny Q\in\Delta_P}\MI(\RV{T}; \RV{Y},\RV{X})\label{eq:bi-syn}\\
			\R{\UI}(\RV{T};\RV{X}_i\backslash\RV{X}_j) 	&= \Min_{\tiny Q\in\Delta_P}\MI(\RV{T};\RV{X}_i,\RV{X}_j) - \Min_{\tiny Q\in\Delta_P}\MI(\RV{T};\RV{X}_j)\label{eq:bi-unq}\quad\text{for all}~\RV{X}_i,\RV{X}_j\in\{\RV{X},\RV{Y}\}\\
			\R{\SI}(\RV{T};\RV{X},\RV{Y}) 	&= \Max_{\tiny Q\in\Delta_P}\CoI(\RV{T};\RV{X};\RV{Y})			
			\label{eq:bi-shr}
		\end{align}\label{eq:bi-pid}
	\end{subequations}
where $\CoI(\RV{T};\RV{X};\RV{Y})$ is the co-information of $\RV{T},\RV{X}$ and $\RV{Y}$ defined as $\MI(\RV{T},\RV{X}) - \MI(\RV{T},\RV{X}\mid\RV{Y})$. Note that~\cite{Chicharro17b} proved that~\eqref{eq:bi-shr} is equivalent to
\begin{equation*}
       \R{\SI}(\RV{T};\RV{X},\RV{Y}) = \Min_{\tiny \substack{Q\in\Delta_P,\\ \CoI(\RV{T};\RV{X};\RV{Y}) = 0}}\MI(\RV{T};\RV{X},\RV{Y})	- 		\Min_{\tiny Q\in\Delta_P}\MI(\RV{T};\RV{X},\RV{Y}).
\end{equation*}
%
	 \subsection{Mutual Information over Bivariate Redundancy Lattice}
		This subsection writes down some mutual information quantities in terms of redundancy lattice partial information measures using~\eqref{eq:lat-main-def}. These formulas will be used in the following subsection to verify that the measures defined in~\eqref{eq:bi-pid} quantify the desired partial information quantities. $\MI(\RV{T};\RV{X},\RV{Y})$ will be the sum of partial information measure on every node of $\mathcal{C}$ as follows:
	 	\begin{equation}
	 		\begin{split}
	 			\MI(\RV{T};\RV{X},\RV{Y})	&= \delta(\RV{T},\RV{X}\RV{Y}) + \delta(\RV{T},\RV{X}) + \delta(\RV{T},\RV{Y}) + \delta(\RV{T}, \RV{X}.\RV{Y}).
	 		\end{split}
	 	\end{equation}
		The mutual information of one source and the target are expressed as
		 \begin{equation}
			 	\begin{split}
			 		\MI(\RV{T};\RV{X}_i)	&= \delta(\RV{T}, \RV{X}_i) + \delta(\RV{T}, \RV{X}_i.\RV{X}_j)\quad\text{for}~\RV{X}_i,\RV{X}_j\in\{\RV{X},\RV{Y}\}.
			 	\end{split}
		 \end{equation}
		The mutual information of one sources and the target conditioned on knowing the other source are expressed as		 
		 \begin{equation}
			 	\begin{split}
			 		\MI(\RV{T};\RV{X}_i\mid\RV{X}_j)	&= \delta(\RV{T},\RV{X}_i\RV{X}_j) + \delta(\RV{T}, \RV{X}_i)\quad\text{for all}~\RV{X}_i,\RV{X}_j\in\{\RV{X},\RV{Y}\}.
			 	\end{split}
		 \end{equation}
		The co-information $\CoI(\RV{T};\RV{X};\RV{Y})$ is expressed as
		 \begin{equation}
		 	\begin{split}
		 		\CoI(\RV{T};\RV{X};\RV{Y})	&= \delta(\RV{T}, \RV{X}.\RV{Y}) - \delta(\RV{T}, \RV{X}\RV{Y}).
		 	\end{split}
		 \end{equation}
	 \subsection{Verification of BROJA Optimization}
		 This subsection will verify that the measures defined in~\eqref{eq:bi-pid} quantify the desired partial information quantities under the maximum decomposition principle. Under assumptions~\ref{assum:bi}, the following statements are valid 
		 \begin{itemize}
		 	\item $\Min_{\tiny Q\in\Delta_P} \delta(\RV{T}, \RV{X}\RV{Y}) = 0$.
		 	\item $\Min_{\tiny Q\in\Delta_P} \delta(\RV{T}, \RV{X}.\RV{Y}) =  \delta(\RV{T}, 1.2), \Min_{\tiny Q\in\Delta_P} \delta(\RV{T}, \RV{X}) =  \delta(\RV{T}, \RV{X})$ and $\Min_{\tiny Q\in\Delta_P} \delta(\RV{T},\RV{Y}) =  \delta(\RV{T},\RV{Y})$.
		 \end{itemize}
		  So, it is easy to see that
		 \begin{equation*}
		 	\begin{split}
		 		\R{\CI}(\RV{T};\RV{X}:\RV{Y})	&= \MI(\RV{T}; \RV{X},\RV{Y}) - \Min_{\tiny Q\in\Delta_P}\MI(\RV{T}; \RV{Y},\RV{X})\\
		 									&= \delta(\RV{T}, \RV{X}\RV{Y})\\
		 		\R{\UI}(\RV{T};\RV{X}\backslash\RV{Y}) 	&= \Min_{\tiny Q\in\Delta_P}\MI(\RV{T};\RV{X},\RV{Y}) - \Min_{\tiny Q\in\Delta_P}\MI(\RV{T};\RV{Y})\\
		 											&= \delta(\RV{T}, \RV{X})\\
		 		\R{\UI}(\RV{T};\RV{Y}\backslash\RV{X}) 	&= \Min_{\tiny Q\in\Delta_P}\MI(\RV{T};\RV{X},\RV{Y}) - \Min_{\tiny Q\in\Delta_P}\MI(\RV{T};\RV{X})\\
		 											&= \delta(\RV{T},\RV{Y}).
		 	\end{split}
		 \end{equation*}
		 Now, $\CoI(\RV{T};\RV{X};\RV{Y}) = 0$ implies that $\delta(\RV{T}, \RV{X}\RV{Y}) = \delta(\RV{T}, \RV{X}.\RV{Y})$, thus
		 \begin{equation*}
		 	\begin{split}
		 		\R{\SI}(\RV{T};\RV{X},\RV{Y}) 	&= \Min_{\tiny \substack{Q\in\Delta_P,\\ \CoI(\RV{T};\RV{X};\RV{Y}) = 0}}\MI(\RV{T};\RV{X},\RV{Y})	- \Min_{\tiny Q\in\Delta_P}\MI(\RV{T};\RV{X},\RV{Y})\\
		 											& = \delta(\RV{T}, \RV{X}.\RV{Y}).
		 	\end{split}
		 \end{equation*}
		 Hence, under assumptions~\ref{assum:bi},
 		\begin{equation}
 			\begin{split}
 				\R{\CI}(\RV{T};\RV{X}:\RV{Y})			&= \CI(\RV{T};\RV{X}:\RV{Y})\\
 				\R{\UI}(\RV{T};\RV{X}\backslash\RV{Y}) 	&= \UI(\RV{T};\RV{X}\backslash\RV{Y})\\
 				\R{\UI}(\RV{T};\RV{Y}\backslash\RV{X}) 	&= \UI(\RV{T};\RV{Y}\backslash\RV{X})\\
 				\R{\SI}(\RV{T};\RV{X},\RV{Y},\RV{Z}) 	&= \SI(\RV{T};\RV{X},\RV{Y},\RV{Z}).
 			\end{split}
 		\end{equation}
		 
\section{Maximum Entropy Decomposition of Trivariate PID}\label{sec:apx-tri-pid}
	Let $\RV{T}$ be the target random variable and $\RV{X},\RV{Y},\RV{Z}$ be the source random variables and $P$ be the joint probability distribution of $(\RV{T},\RV{X},\RV{Y},\RV{Z})$.~\cite{Chicharro17b} using maximum entropy decomposes mutual information $\MI(\RV{T},\RV{X},\RV{Y},\RV{Z})$ into: Synergistic, unique, unique redundant, and redundant information. In this decomposition,
	\begin{itemize}\label{itm:tri-pid}
		\item the synergistic quantity, $\R{\CI}(\RV{T};\RV{X},\RV{Y},\RV{Z})$, captures the sum of all individual synergistic terms, namely, 
		$\delta(\RV{T};\RV{X}\RV{Y}\RV{Z}) + \delta(\RV{T};\RV{X}\RV{Y}) + \delta(\RV{T};\RV{X}\RV{Z}) + \delta(\RV{T};\RV{Y}\RV{Z}) + \delta(\RV{T};\RV{X}\RV{Y}.\RV{X}\RV{Z}) + \delta(\RV{T};\RV{X}\RV{Y}.\RV{Y}\RV{Z}) + \delta(\RV{T};\RV{X}\RV{Z}.\RV{Y}\RV{Z}) + \delta(\RV{T};\RV{X}\RV{Y}.\RV{X}\RV{Z}.\RV{Y}\RV{Z})$,
		\item the unique information, $\R{\UI}(\RV{T};\RV{X}_i\backslash\RV{X}_j,\RV{X}_k)$, captures the sum of the information that $\RV{X}_i$ has about $\RV{T}$ solely, $\delta(\RV{T};\RV{X}_i)$, and the information $\RV{X}_i$ knows redundantly with the synergy of $(\RV{X}_j, \RV{X}_k)$, $\delta(\RV{T};\RV{X}_i.\RV{X}_j\RV{X}_k)$ for all $\RV{X}_i,\RV{X}_j,\RV{X}_k\in\{\RV{X},\RV{Y},\RV{Z}\}$,
		\item the unique redundant information, $\R{\UI}(\RV{T};\RV{X}_i,\RV{X}_j\backslash\RV{X}_k)$, captures the actual unique information that $\RV{X}_i$ and $\RV{X}_j$ has redundantly about $\RV{T}$, $\delta(\RV{T};\RV{X}_i.\RV{X}_j)$ for all $\RV{X}_i,\RV{X}_j,\RV{X}_k\in\{\RV{X},\RV{Y},\RV{Z}\}$, 
		\item and the redundant information, $\R{\SI}(\RV{T};\RV{X},\RV{Y},\RV{Z})$ captures the actual redundant information of $\RV{X},\RV{Y}$ and $\RV{Z}$ about $\RV{T}$, i.e, $\delta(\RV{T};\RV{X}.\RV{Y}.\RV{Z})$.
	\end{itemize}
	Using Beer-Williams axioms the decomposition yields these identities:
	\begin{equation}\label{eq:ids-multi}
		\begin{split}
			\MI(\RV{T};\RV{X},\RV{Y},\RV{Z}) 	&=	\R{\CI}(\RV{T};\RV{X},\RV{Y},\RV{Z}) + \R{\SI}(\RV{T};\RV{X};\RV{Y};\RV{Z})\\
												&+ 	\R{\UI}(\RV{T};\RV{X}\backslash\RV{Y},\RV{Z}) + \R{\UI}(\RV{T};\RV{Y}\backslash\RV{X},\RV{Z}) + \R{\UI}(\RV{T};\RV{Z}\backslash\RV{X},\RV{Y})\\
												&+ 	\R{\UI}(\RV{T};\RV{X},\RV{Y}\backslash\RV{Z}) + \R{\UI}(\RV{T};\RV{X},\RV{Z}\backslash\RV{Y}) + \R{\UI}(\RV{T};\RV{Y},\RV{Z}\backslash\RV{X})\\
			\MI(\RV{T};\RV{X}_i) 					&=	\R{\SI}(\RV{T};\RV{X}_i;\RV{X}_j;\RV{X}_k) + \R{\UI}(\RV{T};\RV{X}_i\backslash\RV{X}_j,\RV{X}_k) +		\R{\UI}(\RV{T};\RV{X}_i,\RV{X}_j\backslash\RV{X}_k)\\
			&+\R{\UI}(\RV{T};\RV{X}_i,\RV{X}_k\backslash\RV{X}_j)\quad\text{for all}~\RV{X}_i,\RV{X}_j,\RV{X}_k\in\{\RV{X},\RV{Y},\RV{Z}\}.
		\end{split}
	\end{equation}
	and $\Delta$ is the set of all joint distributions of $(\RV{T},\RV{X},\RV{Y},\RV{Z})$. 
	The measure uses the maximum entropy decomposition over $\Delta_P$ in order to compute the above quantities. Moreover,~\cite{Chicharro17b} assumes some assumptions over the partial information measures of the redundancy lattice.
	\begin{assum}[Assumption a.1 and Assumption a.2 in~\cite{Chicharro17b}]\label{assm:multi}
	On the trivariate redundancy lattice (Figure~\ref{fig:lattice}), the following assumptions are made to quantify the PID
	\begin{enumerate}
		\item All partial information measures of the redundancy lattice are nonnegative.\label{assm:tri-neg}
		\item The terms $\delta(\RV{T};\RV{X}.\RV{Y}.\RV{Z})$ and $ \delta(\RV{T};\RV{X}_i.\RV{X}_j)$ for all $\RV{X}_i,\RV{X}_j\in\{\RV{X},\RV{Y},\RV{Z}\}$ are invariant on $\Delta_P$.\label{assm:tri-cst} 
		\item The summands $\delta(\RV{T};\RV{X}_i) + \delta(\RV{T};\RV{X}_i.\RV{X}_j\RV{X}_k)$ for all $\RV{X}_i,\RV{X}_j,\RV{X}_k\in\{\RV{X},\RV{Y},\RV{Z}\}$ are invariant on $\Delta_P$.\label{assm:tri-sum-cst}
		\item The terms $\delta(\RV{T};\RV{X}\RV{Y}\RV{Z}), \delta(\RV{T};\RV{X}\RV{Y}.\RV{X}\RV{Z}.\RV{Y}\RV{Z}), \delta(\RV{T};\RV{X}_i\RV{X}_j), \delta(\RV{T};\RV{X}_i\RV{X}_j.\RV{X}_i\RV{X}_k), \delta(\RV{T};\RV{X}_i),$ and $\delta(\RV{T};\RV{X}_i.\RV{X}_j\RV{X}_k)$ for all $\RV{X}_i,\RV{X}_j,\RV{X}_k\in\{\RV{X},\RV{Y},\RV{Z}\}$ are not constant on $\Delta_P$.\label{assm:tri-var}
		\item All synergistic terms, $\delta(\RV{T};\RV{X}\RV{Y}\RV{Z}),$ $\delta(\RV{T};\RV{X}\RV{Y}.\RV{X}\RV{Z}.\RV{Y}\RV{Z}),$ $\delta(\RV{T};\RV{X}_i\RV{X}_j),$ and $\delta(\RV{T};\RV{X}_i\RV{X}_j.\RV{X}_i\RV{X}_k)$  for all $\RV{X}_i,\RV{X}_j,\RV{X}_k\in\{\RV{X},\RV{Y},\RV{Z}\}$ vanishes at the minimum over  $\Delta_P$.\label{assm:tri-syn}
		\item The partial information measures $\delta(\RV{T};\RV{X}_i.\RV{X}_j\RV{X}_k)$ for all $\RV{X}_i,\RV{X}_j,\RV{X}_k\in\{\RV{X},\RV{Y},\RV{Z}\}$ vanishes at the minimum over $\Delta_P$\label{assm:tri-red-syn}.
	\end{enumerate}
	\end{assum}
 	
 	Under the above assumptions and using maximal entropy decomposition,~\cite{Chicharro17b} defines the following optimization problems that compute the PID quantities. 
	 \begin{subequations}
	 	\begin{align}\displaystyle
	 		\R{\CI}(\RV{T};\RV{X},\RV{Y},\RV{Z})			&= \MI(\RV{T}; \RV{X},\RV{Y},\RV{Z}) - \Min_{\tiny Q\in\Delta_P}\MI(\RV{T}; \RV{Y},\RV{X},\RV{Z})\label{eq:multi-syn}\\
	 		\R{\UI}(\RV{T};\RV{X}_i\backslash\RV{X}_j,\RV{X}_k) 	&= \Min_{\tiny Q\in\Delta_P}\MI(\RV{T};\RV{X}_i,\RV{X}_j,\RV{X}_k) - \Min_{\tiny Q\in\Delta_P}\MI(\RV{T};\RV{X}_j,\RV{X}_k)\label{eq:multi-unqX1}\\
	 		 &\text{for all}~\RV{X}_i,\RV{X}_j,\RV{X}_k\in\{\RV{X},\RV{Y},\RV{Z}\}\nonumber\\
	 		\R{\UI}(\RV{T};\RV{X}_i,\RV{X}_j\backslash\RV{X}_k) 	&= \Min_{\tiny \substack{Q\in\Delta_P,\\ \CoI(\RV{T};\RV{X}_i;\RV{X}_j\mid \RV{X}_k) = 0}}\MI(\RV{T};\RV{X}_i,\RV{X}_j,\RV{X}_k) - \Min_{Q\in\Delta_P}\MI(\RV{T};\RV{X}_i,\RV{X}_j,\RV{X}_k)\label{eq:multi-unq-shrX1X2}\\
	 		&\text{for all}~\RV{X}_i,\RV{X}_j,\RV{X}_k\in\{\RV{X},\RV{Y},\RV{Z}\}\nonumber\\
	 		\R{\SI}(\RV{T};\RV{Z},\RV{Y},\RV{Z}) 			&= \Min_{\tiny \substack{Q\in\Delta_P, \CoI(\RV{T};\RV{X};\RV{Y}) = 0,\\ \CoI(\RV{T};\RV{X};\RV{Y}\mid \RV{Z}) = 0,w(Q)}}\MI(\RV{T};\RV{X},\RV{Y},\RV{Z})	- 		\Min_{\tiny \substack{Q\in\Delta_P, w(Q),\\ \CoI(\RV{T};\RV{X};\RV{Y}\mid \RV{Z}) = 0}}\MI(\RV{T};\RV{X},\RV{Y},\RV{Z}),
	 		\label{eq:multi-shr}
	 	\end{align}\label{eq:tri-pid}
	 \end{subequations}
	 where 
	 \begin{equation*}
	 	\begin{aligned}
	 		w(Q):=	\{
	 				 	Q\in \Delta:	&\MI(\RV{T};\RV{X},\RV{Y}) = \Min_{Q\in\Delta_P} \MI(\RV{T};\RV{X},\RV{Y}),\\
	 				 					&\MI(\RV{T};\RV{X},\RV{Z}) = \Min_{Q\in\Delta_P} \MI(\RV{T};\RV{X},\RV{Z}), \\
	 				 					&\MI(\RV{T};\RV{Y},\RV{Z}) = \Min_{Q\in\Delta_P} \MI(\RV{T};\RV{Y},\RV{Z})
	 				\}.
	 	\end{aligned}
	 \end{equation*}
 	\subsection{Mutual Information Over the Trivariate Redundancy Lattice}
        This subsection writes down some mutual information quantities in terms of the trivariate redundancy lattice's partial information measures using~\eqref{eq:lat-main-def}. The verification that the optimization defined in~\eqref{eq:tri-pid} quantifies the desired partial information quantities is discussed in details by~\cite{Chicharro17b} and so will be skipped. But these formulas are needed later when discussing how to compute the individual PID terms using a hierarchy of BROJA and~\cite{Chicharro17b} PID decompositions. The mutual information quantities in terms of redundancy lattice partial information measures. 
 		 
 		$\MI(\RV{T};\RV{X},\RV{Y},\RV{Z})$ will be the sum of partial information measure on every node of the redundancy lattice $\mathcal{C}$ as follows.
 		 \begin{equation}
 			 	\begin{split}
 			 		\MI(\RV{T};\RV{X},\RV{Y},\RV{Z})	&= \delta(\RV{T}, \RV{X}\RV{Y}\RV{Z})+ \delta(\RV{T}, \RV{X}\RV{Y}) + \delta(\RV{T}, \RV{X}\RV{Z}) + \delta(\RV{T}, \RV{Y}\RV{Z}) + \delta(\RV{T}, \RV{X}\RV{Y}.\RV{X}\RV{Z})\\
 			 											&+ \delta(\RV{T}, \RV{X}\RV{Y}.\RV{Y}\RV{Z}) + \delta(\RV{T}, \RV{X}\RV{Z}.\RV{Y}\RV{Z}) + \delta(\RV{T}, \RV{X}\RV{Y}.\RV{X}\RV{Z}.\RV{Y}\RV{Z})\\
 			 											&+ \delta(\RV{T}, \RV{X}) + \delta(\RV{T},\RV{Y}) + \delta(\RV{T},\RV{Z}) + \delta(\RV{T}, \RV{X}.\RV{Y}\RV{Z}) + \delta(\RV{T}, \RV{Y}.\RV{X}\RV{Z}) + \delta(\RV{T}, \RV{Z}.\RV{X}\RV{Y})\\
 			 											&+ \delta(\RV{T}, \RV{X}.\RV{Y}) + \delta(\RV{T}, \RV{X}.\RV{Z}) + \delta(\RV{T}, \RV{Y}.\RV{Z}) + \delta(\RV{T}, \RV{X}.\RV{Y}.\RV{Z}).
 			 											\label{eq:lat-mult-mi}
 			 	\end{split}
 		 \end{equation}
 		 
 		 The mutual information of two sources (jointly) and the target are expressed as
 		 \begin{equation}
 			 	\begin{split}
 			 		\MI(\RV{T};\RV{X}_i,\RV{X}_j)	&= \delta(\RV{T}, \RV{X}_i\RV{X}_j) + \delta(\RV{T}, \RV{X}_i\RV{X}_j.\RV{X}_i\RV{X}_k) + \delta(\RV{T}, \RV{X}_i\RV{X}_j.\RV{X}_j\RV{X}_k)\\
 			 				 					&+ \delta(\RV{T}, \RV{X}_i\RV{X}_j.\RV{X}_i\RV{X}_k.\RV{X}_j\RV{X}_k) + \delta(\RV{T}, \RV{X}_i) + \delta(\RV{T},\RV{X}_j) + \delta(\RV{T}, \RV{X}_i.\RV{X}_j\RV{X}_k)\\
 			 				 					&+ \delta(\RV{T}, \RV{X}_j.\RV{X}_i\RV{X}_k) + \delta(\RV{T}, \RV{X}_k.\RV{X}_i\RV{X}_j) + \delta(\RV{T}, \RV{X}_i.\RV{X}_j) + \delta(\RV{T}, \RV{X}_i.\RV{X}_k)\\
 			 				 					&+ \delta(\RV{T}, \RV{X}_j.\RV{X}_k) + \delta(\RV{T}, \RV{X}_i.\RV{X}_j.\RV{X}_k)\quad\text{for all}~\RV{X}_i,\RV{X}_j,\RV{X}_k\in\{\RV{X},\RV{Y},\RV{Z}\}.
 			 									\label{eq:lat-bi-mi}
 			 	\end{split}
 		 \end{equation}
 		 
 		 The mutual information of one source and the target are expressed as
 		 \begin{equation}
 			 	\begin{split}
 			 		\MI(\RV{T};\RV{X}_i)	&= \delta(\RV{T}, \RV{X}_i) + \delta(\RV{T}, \RV{X}_i.\RV{X}_j\RV{X}_k) + \delta(\RV{T}, \RV{X}_i.\RV{X}_j) + \delta(\RV{T}, \RV{X}_i.\RV{X}_k)\\
 			 		 &+ \delta(\RV{T}, \RV{X}_i.\RV{X}_j.\RV{X}_k)\quad\text{for all}~\RV{X}_i,\RV{X}_j,\RV{X}_k\in\{\RV{X},\RV{Y},\RV{Z}\}.
 			 		\label{eq:lat-sig-mi}
 			 	\end{split}
 		 \end{equation}
 		 
 		 The mutual information of two sources (jointly) and the target conditioned on knowing the other source are expressed as
 		 \begin{equation}
 		 \begin{split}
 			 		\MI(\RV{T};\RV{X}_i,\RV{X}_j\mid\RV{X}_k)	&= \delta(\RV{T}, \RV{X}_i\RV{X}_j\RV{X}_k)+ \delta(\RV{T}, \RV{X}_i\RV{X}_j) + \delta(\RV{T}, \RV{X}_i\RV{X}_k) + \delta(\RV{T},\RV{X}_j\RV{X}_k)\\
 			 				 							&+ \delta(\RV{T}, \RV{X}_i\RV{X}_j.\RV{X}_i\RV{X}_k) + \delta(\RV{T}, \RV{X}_i\RV{X}_j.\RV{X}_j\RV{X}_k) + \delta(\RV{T}, \RV{X}_i\RV{X}_k.\RV{X}_j\RV{X}_k)\\
 			 				 							&+ \delta(\RV{T}, \RV{X}_i\RV{X}_j.\RV{X}_i\RV{X}_k.\RV{X}_j\RV{X}_k) + \delta(\RV{T}, \RV{X}_i) + \delta(\RV{T}, \RV{X}_j)+ \delta(\RV{T}, \RV{X}_i.\RV{X}_j\RV{X}_k)\\ 
 			 				 							&+ \delta(\RV{T}, \RV{X}_j.\RV{X}_i\RV{X}_k) + \delta(\RV{T}, \RV{X}_i.\RV{X}_j)\quad\text{for all}~\RV{X}_i,\RV{X}_j,\RV{X}_k\in\{\RV{X},\RV{Y},\RV{Z}\}.
 			 											\label{eq:lat-bi-cond-mi}
 		 \end{split}
 		 \end{equation} 
 		 
 		 The mutual information of one sources and the target conditioned on knowing only one of the other sources are expressed as		 
 		 \begin{equation}
 			 	\begin{split}
 			 		\MI(\RV{T};\RV{X}_i\mid\RV{X}_j)	&= \delta(\RV{T},\RV{X}_i\RV{X}_j) + \delta(\RV{T}, \RV{X}_i\RV{X}_j.\RV{X}_i\RV{X}_k) + \delta(\RV{T}, \RV{X}_i\RV{X}_j.\RV{X}_j\RV{X}_k)\\
 			 				 						&+ \delta(\RV{T}, \RV{X}_i\RV{X}_j.\RV{X}_i\RV{X}_k.\RV{X}_j\RV{X}_k) + \delta(\RV{T}, \RV{X}_i) + \delta(\RV{T}, \RV{X}_i.\RV{X}_j\RV{X}_k)\\
 			 				 						&+ \delta(\RV{T}, \RV{X}_k.\RV{X}_i\RV{X}_j) + \delta(\RV{T}, \RV{X}_i.\RV{X}_k)\quad\text{for all}~\RV{X}_i,\RV{X}_j,\RV{X}_k\in\{\RV{X},\RV{Y},\RV{Z}\}.
 			 										\label{eq:lat-sig-sig-cond-mi}
 			 	\end{split}
 		 \end{equation}
 		 
 		 The mutual information of one sources and the target conditioned on knowing the other sources are expressed as
 		 \begin{equation}
 			 	\begin{split}
 			 		\MI(\RV{T};\RV{X}_i\mid\RV{X}_j,\RV{X}_k)	&= \delta(\RV{T}, \RV{X}_i\RV{X}_j\RV{X}_k) + \delta(\RV{T}, \RV{X}_i\RV{X}_j) + \delta(\RV{T}, \RV{X}_i\RV{X}_k) + \delta(\RV{T}, \RV{X}_i\RV{X}_j.\RV{X}_i\RV{X}_k)\\ 
 			 		  &+ \delta(\RV{T}, \RV{X}_i)\quad\text{for all}~\RV{X}_i,\RV{X}_j,\RV{X}_k\in\{\RV{X},\RV{Y},\RV{Z}\}.
 			 		\label{eq:lat-sig-bi-cond-mi}
 				\end{split}
 		 \end{equation}		 
 		 
 		 The co-information of two sources and the target are expressed as
 		 \begin{equation}
 			 	\begin{split}
 			 		\CoI(\RV{T};\RV{X}_i;\RV{X}_j)	&= \delta(\RV{T}, \RV{X}_i.\RV{X}_j) + \delta(\RV{T}, \RV{X}_i.\RV{X}_j.\RV{X}_k) - \Bigl(\delta(\RV{T}, \RV{X}_i\RV{X}_j) + \delta(\RV{T}, \RV{X}_i\RV{X}_j.\RV{X}_i\RV{X}_k) \\
 			 									&+ \delta(\RV{T}, \RV{X}_i\RV{X}_j.\RV{X}_j\RV{X}_k) + \delta(\RV{T}, \RV{X}_i\RV{X}_j.\RV{X}_i\RV{X}_k.\RV{X}_j\RV{X}_k) + \delta(\RV{T}, \RV{X}_k.\RV{X}_i\RV{X}_j) \Bigr)\\
 			 									&\text{for all}~\RV{X}_i,\RV{X}_j,\RV{X}_k\in\{\RV{X},\RV{Y},\RV{Z}\}.
 			 									\label{eq:lat-sig-coi}
 			 	\end{split}
 		 \end{equation}
 			
 			The co-information of one sources, two sources (jointly), and the target are expressed as
 			\begin{equation}
 			 	\begin{split}
 			 		\CoI(\RV{T};\RV{X}_i;\RV{X}_j,\RV{X}_k)	&= \delta(\RV{T}, \RV{X}_i.\RV{X}_j\RV{X}_k) + \delta(\RV{T}, \RV{X}_i.\RV{X}_j) + \delta(\RV{T}, \RV{X}_i.\RV{X}_k) + \delta(\RV{T}, \RV{X}_i.\RV{X}_j.\RV{X}_k)\\
 			 											&- \Bigl(\delta(\RV{T}, \RV{X}_i\RV{X}_j\RV{X}_k) + \delta(\RV{T}, \RV{X}_i\RV{X}_j) + \delta(\RV{T}, \RV{X}_i\RV{X}_k) + \delta(\RV{T}, \RV{X}_i\RV{X}_j.\RV{X}_i\RV{X}_k)\Bigr)\\
 			 											&\text{for}~\RV{X}_i,\RV{X}_j,\RV{X}_k\in\{\RV{X},\RV{Y},\RV{Z}\}.
 			 											\label{eq:lat-sig-bi-coi}
 			 	\end{split}
 		 \end{equation}		 

 			The co-information of two sources (jointly), two sources (jointly), and the target are expressed as
 			\begin{equation}
 			 	\begin{split}
 			 		\CoI(\RV{T};\RV{X}_i,\RV{X}_j;\RV{X}_i,\RV{X}_k)	&= \delta(\RV{T}, \RV{X}_i\RV{X}_j.\RV{X}_i\RV{X}_k)  + \delta(\RV{T}, \RV{X}_i\RV{X}_j.\RV{X}_i\RV{X}_k.\RV{X}_j\RV{X}_k)\\
 			 					 				 				&+ \delta(\RV{T}, \RV{X}_i)  + \delta(\RV{T}, \RV{X}_i.\RV{X}_j\RV{X}_k) + \delta(\RV{T}, \RV{X}_j.\RV{X}_i\RV{X}_k) + \delta(\RV{T}, \RV{X}_k.\RV{X}_i\RV{X}_j)\\
 			 					 				 				&+ \delta(\RV{T}, \RV{X}_i.\RV{X}_j) + \delta(\RV{T}, \RV{X}_i.\RV{X}_k) + \delta(\RV{T}, \RV{X}_j.\RV{X}_k) + \delta(\RV{T}, \RV{X}_i.\RV{X}_j.\RV{X}_k)\\
 			 					 				 				&- \delta(\RV{T}, \RV{X}_i\RV{X}_j\RV{X}_k) - \delta(\RV{T}, \RV{X}_j\RV{X}_k)\quad\text{for all}~\RV{X}_i,\RV{X}_j,\RV{X}_k\in\{\RV{X},\RV{Y},\RV{Z}\}.			 					 				 				
 			 					 				 				\label{eq:lat-bi-bi}
 			 	\end{split}
 		 \end{equation}		 
 		 
 		 The co-information of two sources and the target conditioning on knowing the other source are expressed as
 		 \begin{equation}
 			 	\begin{split}
 			 		\CoI(\RV{T};\RV{X}_i;\RV{X}_j\mid\RV{X}_k)	&= \delta(\RV{T}, \RV{X}_i\RV{X}_k.\RV{X}_j\RV{X}_k) + \delta(\RV{T}, \RV{X}_i\RV{X}_j.\RV{X}_i\RV{X}_k.\RV{X}_j\RV{X}_k)\\
 			 				 								&+ \delta(\RV{T}, \RV{X}_i.\RV{X}_j\RV{X}_k) + \delta(\RV{T}, \RV{X}_j.\RV{X}_i\RV{X}_k) + \delta(\RV{T}, \RV{X}_i.\RV{X}_j)\\
 			 				 								&- \delta(\RV{T}, \RV{X}_i\RV{X}_j\RV{X}_k) - \delta(\RV{T}, \RV{X}_i\RV{X}_k)\quad\text{for}~\RV{X}_i,\RV{X}_j,\RV{X}_k\in\{\RV{X},\RV{Y},\RV{Z}\}.			 	
 			 				 								\label{eq:lat-sig-cond-coi}
 			 	\end{split}
 		 \end{equation}
\section{Separating Trivariate PID quantities of Maximum Entropy Decomposition PID}\label{sec:apx-tri-fine}
    In Appendix~\ref{sec:apx-tri-pid}, the maximum entropy decomposition for trivariate PID returns a synergistic term which is the sum of all individual synergy quantities and unique term which the sum of unique and unique redundancy quantities. This section aims to show how to use maximum entropy decomposition for bivariate PID in order to obtain each individual synergy quantity as well as each individual unique and unique redundancy quantity. 
    
	Let $\RV{T}$ be the target random variable and $\RV{X},\RV{Y},\RV{Z}$ be the source random variables and $P$ be the joint probability distribution of $(\RV{T},\RV{X},\RV{Y},\RV{Z})$. Now BROJA will be applied to some subsystems of $(\RV{T},\RV{X},\RV{Y},\RV{Z})$, namely, $(\RV{T},(\RV{X_i},\RV{X_j}),\RV{X_k}),$ (One Singled source) and $(\RV{T},(\RV{X_i},\RV{X_j}),(\RV{X_k},\RV{X_\ell}))$ (Two Double sources) for all $\RV{X_i},\RV{X_j},\RV{X_k},\RV{X_\ell}\in\{\RV{X},\RV{Y},\RV{Z}\}$. Consider the following probability polytopes upon which the optimization will be carried
	\begin{equation}
		\begin{split}
			\Delta_P &= \{Q\in\Delta; Q(\RV{T},\RV{X}) = P(\RV{T},\RV{X}), Q(\RV{T},\RV{Y}) = P(\RV{T},\RV{Y}), Q(\RV{T},\RV{Z}) = P(\RV{T},\RV{Z})\}\\
			\Delta_P^{X_i,X_j.X_k} &= \{Q\in\Delta; Q(\RV{T},\RV{X_i},\RV{X_j}) = P(\RV{T},\RV{X_i},\RV{X_j}), Q(\RV{T},\RV{X_k}) = P(\RV{T},\RV{X_k})\}\\
			&\quad \text{where}~\RV{X_i}\neq\RV{X_j}, \RV{X_i}\neq\RV{X_k}, \RV{X_j}\neq\RV{X_k},\\ 
			&\quad \text{for all}~\RV{X_i},\RV{X_j},\RV{X_k}\in\{\RV{X},\RV{Y},\RV{Z}\}\\
			\Delta_P^{X_i,X_j.X_k,X_\ell} &= \{Q\in\Delta; Q(\RV{T},\RV{X_i},\RV{X_j}) = P(\RV{T},\RV{X_i},\RV{X_j}), Q(\RV{T},\RV{X_k},\RV{X_\ell}) = P(\RV{T},\RV{X_k},\RV{X_\ell})\}\\ 
			&\quad \text{where}~\RV{X_i}\neq\RV{X_j}, \RV{X_k}\neq\RV{X_\ell}, (\RV{X_i},\RV{X_j})\neq(\RV{X_k},\RV{X_\ell})\\
			&\quad\text{for all}~\RV{X_i},\RV{X_j},\RV{X_k},\RV{X_\ell}\in\{\RV{X},\RV{Y},\RV{Z}\}.
		\end{split}
	\end{equation}
	Note that $\Delta_P \subsetneq \Delta_P^{X_i,X_j.X_k}\subsetneq \Delta_P^{X_i,X_j.X_k,X_\ell}$ for all  $\RV{X_i},\RV{X_j},\RV{X_k},\RV{X_\ell}\in\{\RV{X},\RV{Y},\RV{Z}\}$. 	 		
	\subsection{One Singled Source Subsystems}\label{subsec:apx-sig-sys}
		These subsystems has the form $(\RV{T},(\RV{X_i},\RV{X_j}),\RV{X_k})$ where $\RV{X_i},\RV{X_j},\RV{X_k}\in\{\RV{X},\RV{Y},\RV{Z}\}$, $\RV{X_i}\neq\RV{X_j}\neq\RV{X_k},$ and $\RV{X_i}\neq\RV{X_k}$. Now apply the BROJA decomposition to the subsystem $(\RV{T},(\RV{X},\RV{Y}),\RV{Z})$. So its four PID quantities are defined as follows: 
		\begin{equation*}
		 	\begin{split}
		 		\R{\CI}(\RV{T};\RV{X},\RV{Y})	&= \MI(\RV{T};\RV{X},\RV{Y},\RV{Z}) - \Min_{\tiny Q\in\Delta_P^{XY.Z}}\MI(\RV{T};\RV{X},\RV{Y},\RV{Z})\\
		 		\R{\UI}(\RV{T};\RV{X},\RV{Y}\backslash\RV{Z}) 	&= \Min_{\tiny Q\in\Delta_P^{XY.Z}}\MI(\RV{T};\RV{X},\RV{Y},\RV{Z}) - \Min_{\tiny Q\in\Delta_P^{XY.Z}}\MI(\RV{T};\RV{Z})\\
				\R{\UI}(\RV{T};\RV{Z}\backslash\RV{X},\RV{Y}) 	&= \Min_{\tiny Q\in\Delta_P^{XY.Z}}\MI(\RV{T};\RV{X},\RV{Y},\RV{Z}) - \Min_{\tiny Q\in\Delta_P^{XY.Z}}\MI(\RV{T};\RV{X},\RV{Y})\\			 		
				\R{\SI}(\RV{T};\RV{X},\RV{Y},\RV{Z}) 	&= \Min_{\tiny \substack{Q\in\Delta_P^{XY.Z},\\ \CoI(\RV{T};\RV{X},\RV{Y};\RV{Z}) = 0}}\MI(\RV{T};\RV{X},\RV{Y},\RV{Z})	- \Min_{\tiny Q\in\Delta_P^{XY.Z}}\MI(\RV{T};\RV{X},\RV{Y},\RV{Z}).
		 	\end{split}
		\end{equation*}
		Note that the $(\RV{X},\RV{Y})$ marginal distribution is fixed. This implies that the mutual information $\MI(\RV{T};\RV{X},\RV{Y}),$ $\MI(\RV{T};\RV{X}\mid\RV{Y}), \MI(\RV{T};\RV{Y}\mid\RV{X}),$ and $\CoI(\RV{T};\RV{X};\RV{Y})$ are invariant over $\Delta_P^{XY.Z}$. Therefore, the summands $\delta(\RV{T}, \RV{X}\RV{Y}) + \delta(\RV{T}, \RV{X}\RV{Y}.\RV{X}\RV{Z}) + \delta(\RV{T}, \RV{X}\RV{Y}.\RV{Y}\RV{Z}) + \delta(\RV{T}, \RV{X}\RV{Y}.\RV{X}\RV{Z}.\RV{Y}\RV{Z}) + \delta(\RV{T}, \RV{Z}.\RV{X}\RV{Y})$ are fixed. But from assumption~\ref{assm:multi} and the fact that $\RV{X},\RV{Y}$ marginal is fixed, the redundancy $\delta(\RV{T};\RV{Z}.\RV{X}\RV{Y})$ is invariant over $\Delta_P^{XY.Z}$. Thus, in addition to assumption~\ref{assm:tri-cst}, the following partial information measures are invariant over $\Delta_P^{XY.Z}$
			\begin{enumerate}
				\item $\delta(\RV{T};\RV{Z}.\RV{X}\RV{Y})$ since the $(\RV{X},\RV{Y})$ marginal is fixed.
				\item $\delta(\RV{T};\RV{Z})$ since $\MI(\RV{T};\RV{Z})$ and $\delta(\RV{T};\RV{Z}.\RV{X}\RV{Y})$ are invariant over $\Delta_P^{XY.Z}$.
				\item $\delta(\RV{T}, \RV{X}\RV{Y}) + \delta(\RV{T}, \RV{X}\RV{Y}.\RV{X}\RV{Z}) + \delta(\RV{T}, \RV{X}\RV{Y}.\RV{Y}\RV{Z}) + \delta(\RV{T}, \RV{X}\RV{Y}.\RV{X}\RV{Z}.\RV{Y}\RV{Z})$ since $\CoI(\RV{T};\RV{X},\RV{Y})$ and $\delta(\RV{T};\RV{Z}.\RV{X}\RV{Y})$ are invariant over $\Delta_P^{XY.Z}$.
			\end{enumerate}
		Thus, using assumptions~\ref{assm:multi} and definition of $\MI(\RV{T};\RV{X},\RV{Y},\RV{Z})$ over the redundancy lattice,
		\begin{equation*}
			\begin{split}
		 		\Min_{\tiny Q\in\Delta_P^{XY.Z}}\MI(\RV{T};\RV{X},\RV{Y},\RV{Z})	&= \delta(\RV{T}, \RV{X}\RV{Y}) + \delta(\RV{T}, \RV{X}\RV{Y}.\RV{X}\RV{Z}) + \delta(\RV{T}, \RV{X}\RV{Y}.\RV{Y}\RV{Z}) + \delta(\RV{T}, \RV{X}\RV{Y}.\RV{X}\RV{Z}.\RV{Y}\RV{Z})\\
		 																			&+ \delta(\RV{T}, \RV{X}) + \delta(\RV{T}, \RV{Y}) + \delta(\RV{T}, \RV{Z}) + \delta(\RV{T}, \RV{X}.\RV{Y}\RV{Z}) + \delta(\RV{T}, \RV{Y}.\RV{X}\RV{Z})\\
		 																			&+ \delta(\RV{T}, \RV{Z}.\RV{X}\RV{Y}) + \delta(\RV{T}, \RV{X}.\RV{Y}) + \delta(\RV{T}, \RV{X}.\RV{Z}) + \delta(\RV{T}, \RV{Y}.\RV{Z}) + \delta(\RV{T}, \RV{X}.\RV{Y}.\RV{Z}).
			\end{split}
		\end{equation*}
		The synergy is evaluate as
		\begin{equation*}
			\begin{split}
				\R{\CI}(\RV{T};\RV{X},\RV{Y})	&= \MI(\RV{T};\RV{X},\RV{Y},\RV{Z}) - \Min_{\tiny Q\in\Delta_P^{XY.Z}}\MI(\RV{T};\RV{X},\RV{Y},\RV{X})\\
											&= \delta(\RV{T}, \RV{X}\RV{Y}\RV{Z}) + \delta(\RV{T}, \RV{X}\RV{Z}) + \delta(\RV{T}, \RV{Y}\RV{Z}) + \delta(\RV{T}, \RV{X}\RV{Z}.\RV{Y}\RV{Z}).
			\end{split}
		\end{equation*}
		The unique information of $(\RV{X},\RV{Y})$ is evaluated as
		\begin{equation*}
			\begin{split}
				\R{\UI}(\RV{T};\RV{X}\RV{Y}\backslash\RV{Z}) 	&= \Min_{\tiny Q\in\Delta_P^{XY.Z}}\MI(\RV{T};\RV{X},\RV{Y},\RV{Z}) - \Min_{\tiny Q\in\Delta_P^{XY.Z}}\MI(\RV{T};\RV{Z})\\
							 								&= \delta(\RV{T}, \RV{X}\RV{Y}) + \delta(\RV{T}, \RV{X}\RV{Y}.\RV{X}\RV{Z}) + \delta(\RV{T}, \RV{X}\RV{Y}.\RV{Y}\RV{Z}) + \delta(\RV{T}, \RV{X}\RV{Y}.\RV{X}\RV{Z}.\RV{Y}\RV{Z})\\
							 								&+ \delta(\RV{T}, \RV{X}) + \delta(\RV{T}, \RV{Y}) + \delta(\RV{T}, \RV{X}.\RV{Y}\RV{Z}) + \delta(\RV{T}, \RV{Y}.\RV{X}\RV{Z}) + \delta(\RV{T}, \RV{X}.\RV{Y}).
			\end{split}
		\end{equation*}
		The unique information of $\RV{Z}$ is evaluated as 
		\begin{equation*}
			\begin{split}
				\R{\UI}(\RV{T};\RV{Z}\backslash\RV{X}\RV{Y}) 	&= \Min_{\tiny Q\in\Delta_P^{XY.Z}}\MI(\RV{T};\RV{X},\RV{Y},\RV{Z}) - \Min_{\tiny Q\in\Delta_P^{XY.Z}}\MI(\RV{T};\RV{X},\RV{Y})\\
												 			&= \delta(\RV{T}, \RV{Z})
			\end{split}
		\end{equation*}
		When $\CoI(\RV{T};\RV{X},\RV{Y};\RV{Z}) = 0$ then
		\begin{equation*}
		    \begin{split}
		        &\delta(\RV{T}, \RV{Z}.\RV{X}\RV{Y}) + \delta(\RV{T}, \RV{X}.\RV{Z}) + \delta(\RV{T}, \RV{Y}.\RV{Z}) + \delta(\RV{T}, \RV{X}.\RV{Y}.\RV{Z})\\ 
		        &=\delta(\RV{T}, \RV{X}\RV{Y}\RV{Z}) + \delta(\RV{T}, \RV{X}\RV{Z}) + \delta(\RV{T}, \RV{Y}\RV{Z})+ \delta(\RV{T}, \RV{X}\RV{Z}.\RV{Y}\RV{Z}).
		    \end{split}
		\end{equation*}
		The shared information of $(\RV{X},\RV{Y})$ and $\RV{Z}$ is evaluated as 
		\begin{equation*}
			\begin{split}
				\R{\SI}(\RV{T};\RV{X},\RV{Y},\RV{Z}) 	&= \Min_{\tiny \substack{Q\in\Delta_P^{XY.Z},\\ \CoI(\RV{T};\RV{X},\RV{Y};\RV{Z}) = 0}}\MI(\RV{T};\RV{X},\RV{Y},\RV{Z})	- \Min_{\tiny Q\in\Delta_P^{XY.Z}}\MI(\RV{T};\RV{X},\RV{Y},\RV{Z})\\	
													&= \delta(\RV{T}, \RV{Z}.\RV{X}\RV{Y}) + \delta(\RV{T}, \RV{X}.\RV{Z}) + \delta(\RV{T}, \RV{Y}.\RV{Z}) + \delta(\RV{T}, \RV{X}.\RV{Y}.\RV{Z}).
			\end{split}
		\end{equation*}
		Hence the BROJA decomposition of the subsystem  $(\RV{T},(\RV{X},\RV{Y}),\RV{Z})$ is 
		\begin{equation*}
		 	\begin{split}
		 		\R{\CI}(\RV{T};(\RV{X},\RV{Y}),\RV{Z})				&= \delta(\RV{T};\RV{X}\RV{Y}\RV{Z}) + \delta(\RV{T};\RV{X}\RV{Z}) + \delta(\RV{T};\RV{Y}\RV{Z}) + \delta(\RV{T};\RV{X}\RV{Z}.\RV{Y}\RV{Z})\\
		 		\R{\UI}(\RV{T};(\RV{X},\RV{Y})\backslash\RV{Z}) 	&= \delta(\RV{T};\RV{X}) + \delta(\RV{T};\RV{X}.\RV{Y}\RV{Z}) + \delta(\RV{T};\RV{Y}) + \delta(\RV{T};\RV{Y}.\RV{X}\RV{Z}) + \delta(\RV{T};\RV{X}.\RV{Y})\\
		 															&+ \delta(\RV{T};\RV{X}\RV{Y}) + \delta(\RV{T};\RV{X}\RV{Y}.\RV{X}\RV{Z}) + \delta(\RV{T};\RV{X}\RV{Y}.\RV{Y}\RV{Z})+ \delta(\RV{T};\RV{X}\RV{Y}.\RV{X}\RV{Z}.\RV{Y}\RV{Z})\\
		 		\R{\UI}(\RV{T};\RV{Z}\backslash\RV{X},\RV{Y}) 		&= \delta(\RV{T};\RV{Z})\\
				\R{\SI}(\RV{T};\RV{X},\RV{Y},\RV{Z}) 				&= \delta(\RV{T};\RV{Z}.\RV{X}\RV{Y}) +\delta(\RV{T};\RV{X}.\RV{Z})+\delta(\RV{T};\RV{Y}.\RV{Z}) +\delta(\RV{T};\RV{X}.\RV{Y}.\RV{Z}).
		 	\end{split}
		\end{equation*}
		Whence the BROJA decompositions of the subsystems  $(\RV{T},(\RV{X},\RV{Z}),\RV{Y})$ and $(\RV{T},(\RV{Y},\RV{Z}),\RV{X})$ are
		\begin{equation*}
		 	\begin{split}
		 		\R{\CI}(\RV{T};(\RV{X},\RV{Z}),\RV{Y})				&= \delta(\RV{T};\RV{X}\RV{Y}\RV{Z}) + \delta(\RV{T};\RV{X}\RV{Y}) + \delta(\RV{T};\RV{Y}\RV{Z}) + \delta(\RV{T};\RV{X}\RV{Y}.\RV{Y}\RV{Z})\\
		 		\R{\UI}(\RV{T};(\RV{X},\RV{Z})\backslash\RV{Y}) 	&= \delta(\RV{T};\RV{X}) + \delta(\RV{T};\RV{X}.\RV{Y}\RV{Z}) + \delta(\RV{T};\RV{Z}) + \delta(\RV{T};\RV{Z}.\RV{X}\RV{Y}) + \delta(\RV{T};\RV{X}.\RV{Z})\\
		 		                                                    &+ \delta(\RV{T};\RV{X}\RV{Z}) + \delta(\RV{T};\RV{X}\RV{Y}.\RV{X}\RV{Z}) + \delta(\RV{T};\RV{X}\RV{Z}.\RV{Y}\RV{Z})+ \delta(\RV{T};\RV{X}\RV{Y}.\RV{X}\RV{Z}.\RV{Y}\RV{Z})\\
		 		\R{\UI}(\RV{T};\RV{Y}\backslash\RV{X},\RV{Z}) 		&= \delta(\RV{T};\RV{Y})\\
				\R{\SI}(\RV{T};\RV{X},\RV{Y},\RV{Z}) 				&= \delta(\RV{T};\RV{Y}.\RV{X}\RV{Z}) +\delta(\RV{T};\RV{X}.\RV{Y})+\delta(\RV{T};\RV{Y}.\RV{Z}) +\delta(\RV{T};\RV{X}.\RV{Y}.\RV{Z}).
		 	\end{split}
		\end{equation*}
		and 
		\begin{equation*}
		 	\begin{split}
		 		\R{\CI}(\RV{T};(\RV{Y},\RV{Z}),\RV{X})				&= \delta(\RV{T};\RV{X}\RV{Y}\RV{Z}) + \delta(\RV{T};\RV{X}\RV{Y}) + \delta(\RV{T};\RV{X}\RV{Z}) + \delta(\RV{T};\RV{X}\RV{Y}.\RV{X}\RV{Z})\\
		 		\R{\UI}(\RV{T};(\RV{Y},\RV{Z})\backslash\RV{X}) 	&= \delta(\RV{T};\RV{Y}) + \delta(\RV{T};\RV{Y}.\RV{X}\RV{Z}) + \delta(\RV{T};\RV{Z}) + \delta(\RV{T};\RV{Z}.\RV{X}\RV{Y}) + \delta(\RV{T};\RV{Y}.\RV{Z})\\
		 		                                                    &+ \delta(\RV{T};\RV{Y}\RV{Z}) + \delta(\RV{T};\RV{X}\RV{Y}.\RV{Y}\RV{Z}) + \delta(\RV{T};\RV{X}\RV{Z}.\RV{Y}\RV{Z})+ \delta(\RV{T};\RV{X}\RV{Y}.\RV{X}\RV{Z}.\RV{Y}\RV{Z})\\
		 		\R{\UI}(\RV{T};\RV{X}\backslash\RV{Y},\RV{Z}) 		&= \delta(\RV{T};\RV{X})\\
				\R{\SI}(\RV{T};\RV{X},\RV{Y},\RV{Z}) 				&= \delta(\RV{T};\RV{X}.\RV{Y}\RV{Z}) +\delta(\RV{T};\RV{X}.\RV{Y})+\delta(\RV{T};\RV{X}.\RV{Z}) +\delta(\RV{T};\RV{X}.\RV{Y}.\RV{Z}).
		 	\end{split}
		\end{equation*}
	\subsection{Two Double Sources Subsystems}\label{subsec:apx-dou-sys}
		These subsystems has the form $(\RV{T},(\RV{X_i},\RV{X_j}),(\RV{X_k},\RV{X_\ell}))$ where $\RV{X_i},\RV{X_j},\RV{X_k},\RV{X_\ell}\in\{\RV{X},\RV{Y},\RV{Z}\},$ $\RV{X_i}\neq\RV{X_j},$ $\RV{X_k}\neq\RV{X_\ell},$ and $(\RV{X_i},\RV{X_j})\neq(\RV{X_k},\RV{X_\ell})$. Now apply the BROJA decomposition to the subsystem $(\RV{T},(\RV{X},\RV{Y}),(\RV{X},\RV{Z}))$. So its four PID quantities are defined as follows 
		\begin{equation*}
		 	\begin{split}
		 		\R{\CI}(\RV{T};\RV{X},\RV{Y})
		 		&= \MI(\RV{T};\RV{X},\RV{Y},\RV{Z}) - \Min_{\tiny Q\in\Delta_P^{XY.XZ}}\MI(\RV{T};\RV{X},\RV{Y},\RV{Z})\\
		 		\R{\UI}(\RV{T};(\RV{X},\RV{Y})\backslash(\RV{X},\RV{Z})) 
		 		&= \Min_{\tiny Q\in\Delta_P^{XY.XZ}}\MI(\RV{T};\RV{X},\RV{Y},\RV{Z}) - \Min_{\tiny Q\in\Delta_P^{XY.XZ}}\MI(\RV{T};\RV{X},\RV{Z})\\
		 		\R{\UI}(\RV{T};(\RV{X},\RV{Z})\backslash(\RV{X},\RV{Y})) 
		 		&= \Min_{\tiny Q\in\Delta_P^{XY.XZ}}\MI(\RV{T};\RV{X},\RV{Y},\RV{Z}) - \Min_{\tiny Q\in\Delta_P^{XY.XZ}}\MI(\RV{T};\RV{X},\RV{Y})\\
				\R{\SI}(\RV{T};(\RV{X},\RV{Y}),(\RV{X},\RV{Z})) 
				&= \Min_{\tiny \substack{Q\in\Delta_P^{XY.XZ},\\ \CoI(\RV{T};\RV{X},\RV{Y};\RV{X},\RV{Z}) = 0}}\MI(\RV{T};\RV{X},\RV{Y},\RV{Z})	- \Min_{\tiny Q\in\Delta_P^{XY.XZ}}\MI(\RV{T};\RV{X},\RV{Y},\RV{Z}).
		 	\end{split}
		\end{equation*}
		Note that the $(\RV{X},\RV{Y})$ and $(\RV{X},\RV{Z})$ marginal distributions are fixed. Then, $\MI(\RV{T};\RV{X_1},\RV{X_2}),$ $\MI(\RV{T};\RV{X_1}\mid\RV{X_2}),$ $\CoI(\RV{T};\RV{X_1};\RV{X_2}),$ and $\MI(\RV{T};\RV{X}:\RV{Y},\RV{X}:\RV{Z})$ are invariant over $\Delta_P^{XY.XZ}$ for $(\RV{X_1},\RV{X_2}) = (\RV{X},\RV{Y})$ and $(\RV{X_1},\RV{X_2}) = (\RV{X},\RV{Z})$. Therefore, $\delta(\RV{T},\RV{X}\RV{Y}) + \delta(\RV{T}, \RV{X}\RV{Y}.\RV{X}\RV{Z}) + \delta(\RV{T}, \RV{X}\RV{Y}.\RV{Y},\RV{Z}) + \delta(\RV{T}, \RV{X}\RV{Y}.\RV{X}\RV{Z}.\RV{Y}\RV{Z}) + \delta(\RV{T}, \RV{Z}.\RV{X}\RV{Y})$ and $\delta(\RV{T},\RV{X},\RV{Z}) + \delta(\RV{T}, \RV{X}\RV{Y}.\RV{X}\RV{Z}) + \delta(\RV{T}, \RV{X}\RV{Z}.\RV{Y}\RV{Z}) + \delta(\RV{T}, \RV{X}\RV{Y}.\RV{X}\RV{Z}.\RV{Y}\RV{Z}) + \delta(\RV{T}, \RV{Y}.\RV{X}\RV{Z})$ are fixed. But from the assumption~\ref{assm:multi} and the two fixed $(\RV{X},\RV{Y})$ and $(\RV{X},\RV{Z})$ marginals, then the redundancies $\delta(\RV{T};\RV{Z}.\RV{X}\RV{Y})$ and $\delta(\RV{T};\RV{Y}.\RV{X}\RV{Z})$ are invariant over $\Delta_P^{XY.XZ}$. Therefore, in addition to assumption~\ref{assm:tri-cst}, the following partial information measures are invariant $\Delta_P^{XY.XZ}$
		\begin{enumerate}
			\item $\delta(\RV{T};\RV{Z}.\RV{X}\RV{Y})$ since $(\RV{X},\RV{Y})$ marginal is fixed.
			\item $\delta(\RV{T};\RV{Y}.\RV{X}\RV{Z})$ since $(\RV{X},\RV{Z})$ marginal is fixed.
			\item $\delta(\RV{T};\RV{X}\RV{Y}.\RV{X}\RV{Z}) + \delta(\RV{T}, \RV{X}\RV{Y}.\RV{X}\RV{Z}.\RV{Y}\RV{Z})$ since $\MI(\RV{T};\RV{X}:\RV{Y},\RV{X}:\RV{Z})$ and $\delta(\RV{T};\RV{Z}.\RV{X}\RV{Y})$ are invariant.
			\item $\delta(\RV{T};\RV{Z})$ since $\MI(\RV{T};\RV{Z})$ and $\delta(\RV{T};\RV{Z}.\RV{X}\RV{Y})$ are invariant over $\Delta_P^{XY.XZ}.$
			\item $\delta(\RV{T};\RV{Y})$ since $\MI(\RV{T};\RV{Y})$ and $\delta(\RV{T};\RV{Y}.\RV{X}\RV{Z})$ are invariant over $\Delta_P^{XY.XZ}.$
			\item $\delta(\RV{T},\RV{X}\RV{Y})+ \delta(\RV{T}, \RV{X}\RV{Y}.\RV{Y}\RV{Z})$ since  $\delta(\RV{T};\RV{X}\RV{Y}.\RV{X}\RV{Z}) + \delta(\RV{T}, \RV{X}\RV{Y}.\RV{X}\RV{Z}.\RV{Y}\RV{Z}),$ and  $\CoI(\RV{T};\RV{X};\RV{Y}),$ $\delta(\RV{T};\RV{Z}.\RV{X}\RV{Y})$ are invariant over $\Delta_P^{XY.XZ}.$
			\item $\delta(\RV{T},\RV{X}\RV{Z}) + \delta(\RV{T}, \RV{X}\RV{Z}.\RV{Y}\RV{Z})$  since $\delta(\RV{T};\RV{X}\RV{Y}.\RV{X}\RV{Z}) + \delta(\RV{T}, \RV{X}\RV{Y}.\RV{X}\RV{Z}.\RV{Y}\RV{Z}),$ $\CoI(\RV{T};\RV{X};\RV{Z}),$ and $\delta(\RV{T};\RV{Y}.\RV{X}\RV{Z})$ are invariant over $\Delta_P^{XY.XZ}.$
		\end{enumerate}
		Thus, using assumptions~\ref{assm:multi} and definition of $\MI(\RV{T};\RV{X},\RV{Y},\RV{Z})$ over the redundancy lattice,
		\begin{equation*}
			\begin{split}
		 		\Min_{\tiny Q\in\Delta_P^{XY.XZ}}\MI(\RV{T};\RV{X},\RV{Y},\RV{Z})	&= \delta(\RV{T}, \RV{X}\RV{Y}) + \delta(\RV{T}, \RV{X}\RV{Z}) + \delta(\RV{T}, \RV{X}\RV{Y}.\RV{X}\RV{Z}) + \delta(\RV{T}, \RV{X}\RV{Y}.\RV{Y}\RV{Z})\\
		 																			&+\delta(\RV{T}, \RV{X}\RV{Z}.\RV{Y}\RV{Z}) + \delta(\RV{T}, \RV{X}\RV{Y}.\RV{X}\RV{Z}.\RV{Y}\RV{Z}) + \delta(\RV{T}, \RV{X}) + \delta(\RV{T},\RV{Y}) \\
		 																			&+ \delta(\RV{T}, \RV{Z}) + \delta(\RV{T}, \RV{X}.\RV{Y}\RV{Z})  + \delta(\RV{T}, \RV{Y}.\RV{X}\RV{Z}) + \delta(\RV{T}, \RV{Z}.\RV{X}\RV{Y})\\ 
		 																			&+ \delta(\RV{T}, \RV{X}.\RV{Y}) + \delta(\RV{T}, \RV{X}.\RV{Z}) + \delta(\RV{T}, \RV{Y}.\RV{Z}) + \delta(\RV{T}, \RV{X}.\RV{Y}.\RV{Z}).
			\end{split}
		\end{equation*}
		The synergy is evaluate as
		\begin{equation*}
			\begin{split}
				\R{\CI}(\RV{T};\RV{X},\RV{Y})	&= \MI(\RV{T};\RV{X},\RV{Y},\RV{Z}) - \Min_{\tiny Q\in\Delta_P^{XY.XZ}}\MI(\RV{T};\RV{X},\RV{Y},\RV{Z})\\
												&=\delta(\RV{T}, \RV{X}\RV{Y}\RV{Z}) + \delta(\RV{T},\RV{Y}\RV{Z}).
			\end{split}
		\end{equation*}
		The unique information of $(\RV{X},\RV{Y})$ is evaluated as
		\begin{equation*}
			\begin{split}
				\R{\UI}(\RV{T};\RV{X},\RV{Y}\backslash\RV{X},\RV{Z}) 	&= \Min_{\tiny Q\in\Delta_P^{XY.XZ}}\MI(\RV{T};\RV{X},\RV{Y},\RV{Z}) - \Min_{\tiny Q\in\Delta_P^{XY.XZ}}\MI(\RV{T};\RV{X},\RV{Z})\\
										 								&= \delta(\RV{T}, \RV{X}\RV{Y}) + \delta(\RV{T}, \RV{X}\RV{Y}.\RV{Y}\RV{Z}) + \delta(\RV{T},\RV{Y}).
			\end{split}
		\end{equation*}
		
		The unique information of $(\RV{X},\RV{Z})$ is evaluated as 
		\begin{equation*}
			\begin{split}
				\R{\UI}(\RV{T};\RV{X},\RV{Z}\backslash\RV{X},\RV{Y}) 	&= \Min_{\tiny Q\in\Delta_P^{XY.XZ}}\MI(\RV{T};\RV{X},\RV{Y},\RV{Z}) - \Min_{\tiny Q\in\Delta_P^{XY.XZ}}\MI(\RV{T};\RV{X},\RV{Y})\\
															 			&= \delta(\RV{T},\RV{Y}\RV{Z}) + \delta(\RV{T}, \RV{X}\RV{Z}.\RV{Y}\RV{Z}) + \delta(\RV{T},\RV{Z}). 
			\end{split}
		\end{equation*}
		When $\CoI(\RV{T};\RV{X},\RV{Y};\RV{X},\RV{Z}) = 0$ then
		\begin{equation*}
			\begin{split}	
			\delta(\RV{T}, \RV{X}\RV{Y}\RV{Z}) + \delta(\RV{T}, \RV{Y}\RV{Z})	&= \delta(\RV{T}, \RV{X}\RV{Y}.\RV{X}\RV{Z}) + \delta(\RV{T}, \RV{X}\RV{Y}.\RV{X}\RV{Z}.\RV{Y}\RV{Z})\\ 
														&+ \delta(\RV{T}, \RV{X})  + \delta(\RV{T}, \RV{X}.\RV{Y}\RV{Z}) + \delta(\RV{T}, \RV{Y}.\RV{X}\RV{Z}) + \delta(\RV{T}, \RV{Z}.\RV{X}\RV{Y})\\
											 			&+ \delta(\RV{T}, \RV{X}.\RV{Y}) + \delta(\RV{T}, \RV{X}.\RV{Z}) + \delta(\RV{T}, \RV{Y}.\RV{Z}) + \delta(\RV{T}, \RV{X}.\RV{Y}.\RV{Z}).
			\end{split}
		\end{equation*}
		The shared information of $\RV{X},\RV{Y}$ and $\RV{X},\RV{Z}$ is evaluated as 
		\begin{equation*}
			\begin{split}
				\R{\SI}(\RV{T};\RV{X},\RV{Y},\RV{Z}) 	&= \Min_{\tiny \substack{Q\in\Delta_P^{XY.XZ},\\ \CoI(\RV{T};\RV{X},\RV{Y};\RV{X},\RV{Z}) = 0}}\MI(\RV{T};\RV{X},\RV{Y},\RV{Z})	- \Min_{\tiny Q\in\Delta_P^{XY.XZ}}\MI(\RV{T};\RV{X},\RV{Y},\RV{Z})\\	
													&= \delta(\RV{T}, \RV{X}\RV{Y}.\RV{X}\RV{Z}) + \delta(\RV{T}, \RV{X}\RV{Y}.\RV{X}\RV{Z}.\RV{Y}\RV{Z})\\ 
														&+ \delta(\RV{T}, \RV{X})  + \delta(\RV{T}, \RV{X}.\RV{Y}\RV{Z}) + \delta(\RV{T}, \RV{Y}.\RV{X}\RV{Z}) + \delta(\RV{T}, \RV{Z}.\RV{X}\RV{Y})\\
											 			&+ \delta(\RV{T}, \RV{X}.\RV{Y}) + \delta(\RV{T}, \RV{X}.\RV{Z}) + \delta(\RV{T}, \RV{Y}.\RV{Z}) + \delta(\RV{T}, \RV{X}.\RV{Y}.\RV{Z}). 
			\end{split}
		\end{equation*}
		Hence then BROJA decomposition of the subsystem  $(\RV{T},(\RV{X},\RV{Y}),(\RV{X},\RV{Z}))$ is 
		\begin{equation*}
		 	\begin{split}
		 		\R{\CI}(\RV{T};(\RV{X},\RV{Y}),(\RV{X},\RV{Z}))				&= \delta(\RV{T};\RV{X}\RV{Y}\RV{Z}) + \delta(\RV{T};\RV{Y}\RV{Z}),\\
		 		\R{\UI}(\RV{T};(\RV{X},\RV{Y})\backslash(\RV{X},\RV{Z})) 	&= \delta(\RV{T};\RV{X}\RV{Y}) + \delta(\RV{T};\RV{X}\RV{Y}.\RV{Y}\RV{Z}) + \delta(\RV{T};\RV{Y}),\\
		 		\R{\UI}(\RV{T};(\RV{X},\RV{Y})\backslash(\RV{X},\RV{Z})) 	&=\delta(\RV{T};\RV{X}\RV{Z}) + \delta(\RV{T};\RV{X}\RV{Z}.\RV{Y}\RV{Z}) + \delta(\RV{T};\RV{Z}),\\
				\R{\SI}(\RV{T};(\RV{X},\RV{Y}),(\RV{X},\RV{Z})) 			&= \delta(\RV{T};\RV{X}\RV{Y}.\RV{X}\RV{Z})\\
																			&+ \delta(\RV{T};\RV{X}) + \delta(\RV{T};\RV{X}\RV{Y}.\RV{X}\RV{Z}.\RV{Y}\RV{Z})\\
																			&+ \delta(\RV{T};\RV{X}.\RV{Y}\RV{Z}) + \delta(\RV{T};\RV{Y}.\RV{X}\RV{Z}) + \delta(\RV{T};\RV{Z}.\RV{X}\RV{Y})\\ 			
																			&+ \delta(\RV{T};\RV{X}.\RV{Y}) + \delta(\RV{T};\RV{X}.\RV{Z})+\delta(\RV{T};\RV{Y}.\RV{Z})\\ 
																			&+ \delta(\RV{T};\RV{X}.\RV{Y}.\RV{Z}).
		 	\end{split}
		\end{equation*}
		Whence the BROJA decomposition of the subsystem $(\RV{T},(\RV{X},\RV{Y}),(\RV{Y},\RV{Z}))$ is 
		\begin{equation*}
		 	\begin{split}
		 		\R{\CI}(\RV{T};(\RV{X},\RV{Y}),(\RV{Y},\RV{Z}))				&= \delta(\RV{T};\RV{X}\RV{Y}\RV{Z}) + \delta(\RV{T};\RV{X}\RV{Z}),\\
		 		\R{\UI}(\RV{T};(\RV{X},\RV{Y})\backslash(\RV{Y},\RV{Z})) 	&= \delta(\RV{T};\RV{X}\RV{Y}) + \delta(\RV{T};\RV{X}\RV{Y}.\RV{X}\RV{Z}) + \delta(\RV{T};\RV{X}),\\
		 		\R{\UI}(\RV{T};(\RV{X},\RV{Y})\backslash(\RV{Y},\RV{Z})) 	&=\delta(\RV{T};\RV{Y}\RV{Z}) + \delta(\RV{T};\RV{X}\RV{Z}.\RV{Y}\RV{Z}) + \delta(\RV{T};\RV{Z}),\\
				\R{\SI}(\RV{T};(\RV{X},\RV{Y}),(\RV{Y},\RV{Z})) 			&= \delta(\RV{T};\RV{X}\RV{Y}.\RV{Y}\RV{Z})\\
																			&+ \delta(\RV{T};\RV{Y}) + \delta(\RV{T};\RV{X}\RV{Y}.\RV{X}\RV{Z}.\RV{Y}\RV{Z})\\
																			&+ \delta(\RV{T};\RV{X}.\RV{Y}\RV{Z}) + \delta(\RV{T};\RV{Y}.\RV{X}\RV{Z}) + \delta(\RV{T};\RV{Z}.\RV{X}\RV{Y})\\ 			
																			&+ \delta(\RV{T};\RV{X}.\RV{Y}) + \delta(\RV{T};\RV{X}.\RV{Z})+\delta(\RV{T};\RV{Y}.\RV{Z})\\ 
																			&+ \delta(\RV{T};\RV{X}.\RV{Y}.\RV{Z}).
		 	\end{split}
		\end{equation*}
		and that of $(\RV{T},(\RV{X},\RV{Z}),(\RV{Y},\RV{Z}))$ is 
		\begin{equation*}
		 	\begin{split}
		 		\R{\CI}(\RV{T};(\RV{X},\RV{Z}),(\RV{Y},\RV{Z}))				&= \delta(\RV{T};\RV{X}\RV{Y}\RV{Z}) + \delta(\RV{T};\RV{X}\RV{Y}),\\
		 		\R{\UI}(\RV{T};(\RV{X},\RV{Z})\backslash(\RV{Y},\RV{Z})) 	&= \delta(\RV{T};\RV{X}\RV{Z}) + \delta(\RV{T};\RV{X}\RV{Y}.\RV{X}\RV{Z}) + \delta(\RV{T};\RV{X}),\\
		 		\R{\UI}(\RV{T};(\RV{Y},\RV{Z})\backslash(\RV{X},\RV{Z})) 	&=\delta(\RV{T};\RV{Y}\RV{Z}) + \delta(\RV{T};\RV{X}\RV{Y}.\RV{Y}\RV{Z}) + \delta(\RV{T};\RV{Y}),\\
				\R{\SI}(\RV{T};(\RV{X},\RV{Z}),(\RV{Y},\RV{Z})) 			&= \delta(\RV{T};\RV{X}\RV{Z}.\RV{Y}\RV{Z})\\
																			&+ \delta(\RV{T};\RV{Z}) + \delta(\RV{T};\RV{X}\RV{Y}.\RV{X}\RV{Z}.\RV{Y}\RV{Z})\\
																			&+ \delta(\RV{T};\RV{X}.\RV{Y}\RV{Z}) + \delta(\RV{T};\RV{Y}.\RV{X}\RV{Z}) + \delta(\RV{T};\RV{Z}.\RV{X}\RV{Y})\\ 			
																			&+ \delta(\RV{T};\RV{X}.\RV{Y}) + \delta(\RV{T};\RV{X}.\RV{Z})+\UI(\RV{T};\RV{Y}.\RV{Z})\\ 
																			&+ \delta(\RV{T};\RV{X}.\RV{Y}.\RV{Z}).
		 	\end{split}
		\end{equation*}
	\subsection{Synergy of Three Double Sources System}\label{subsec:apx-trio-sys}
		Consider the system of the form $(\RV{T},(\RV{X},\RV{Y}),(\RV{X},\RV{Z}),(\RV{Y},\RV{Z}))$. The sources here are called composite since they are compositions of the primary sources $\RV{X},\RV{Y}$ and $\RV{Z}$.~\cite{Chicharro17b} measures using maximum entropy decomposition~\eqref{eq:tri-pid} can be used to capture the synergy of composite sources but do not break down contributions that involve unique or redundancy of composite sources. The optimization is taken over the polytope:
		\begin{equation}
			\begin{split}
				\Delta_P^{XY.XZ.YZ} = \{Q\in\Delta;	&\quad Q(\RV{T},\RV{X},\RV{Y}) = P(\RV{T},\RV{X},\RV{Y}),\\ 
													&\quad Q(\RV{T},\RV{X},\RV{Z}) = P(\RV{T},\RV{X},\RV{Z}),\\ 
													&\quad Q(\RV{T},\RV{Y},\RV{Z}) = P(\RV{T},\RV{Y},\RV{Z})\}. 
			\end{split}
		\end{equation}
		In this polytope, $\MI(\RV{X_i},\RV{X_j}),\CoI(\RV{X_i},\RV{X_j}),$ and  $\MI(\RV{X_i}\mid\RV{X_j})$ are invariant for all $\RV{X_i},\RV{X_j}\in\{\RV{X},\RV{Y},\RV{Z}\}$. Therefore, in addition to assumption~\ref{assm:tri-cst}, the following partial information measures are invariant $\Delta_P^{XY.XZ.YZ}$
		\begin{enumerate}
			\item $\delta(\RV{T};\RV{Z}.\RV{X}\RV{Y})$ since $(\RV{X},\RV{Y})$ marginal is fixed.
			\item $\delta(\RV{T};\RV{Y}.\RV{X}\RV{Z})$ since $(\RV{X},\RV{Z})$ marginal is fixed.
			\item $\delta(\RV{T};\RV{X}.\RV{Y}\RV{Z})$ since $(\RV{Y},\RV{Z})$ marginal is fixed.
			\item $\delta(\RV{T};\RV{X}\RV{Y}.\RV{X}\RV{Z})$ since $(\RV{X},\RV{Y}),$ $(\RV{X},\RV{Z}),$ and $(\RV{Y},\RV{Z})$ marginals are fixed.
			\item $\delta(\RV{T};\RV{X}\RV{Y}.\RV{Y}\RV{Z})$ since $(\RV{X},\RV{Y}),$ $(\RV{X},\RV{Z}),$ and $(\RV{Y},\RV{Z})$ marginals are fixed.
			\item $\delta(\RV{T};\RV{X}\RV{Z}.\RV{Y}\RV{Z})$ since $(\RV{X},\RV{Y}),$ $(\RV{X},\RV{Z}),$ and $(\RV{Y},\RV{Z})$ marginals are fixed.
			\item $\delta(\RV{T};\RV{X}\RV{Y}.\RV{X}\RV{Z}.\RV{Y}\RV{Z})$ since $(\RV{X},\RV{Y}),$ $(\RV{X},\RV{Z})$ and $(\RV{Y},\RV{Z})$ marginals are fixed.
			\item $\delta(\RV{T};\RV{Z})$ since $\MI(\RV{T};\RV{Z})$ and $\delta(\RV{T};\RV{Z}.\RV{X}\RV{Y})$ are invariant over $\Delta_P^{XY.XZ.YZ}.$
			\item $\delta(\RV{T};\RV{Y})$ since $\MI(\RV{T};\RV{Y})$ and $\delta(\RV{T};\RV{Y}.\RV{X}\RV{Z})$ are invariant over $\Delta_P^{XY.XZ.YZ}.$
			\item $\delta(\RV{T};\RV{X})$ since $\MI(\RV{T};\RV{X})$ and $\delta(\RV{T};\RV{X}.\RV{Y}\RV{Z})$ are invariant over $\Delta_P^{XY.XZ.YZ}.$
			\item $\delta(\RV{T}, \RV{X}\RV{Y})$ since  $\CoI(\RV{T};\RV{X};\RV{Y})$, $\delta(\RV{T},\RV{X}\RV{Y}.\RV{X}\RV{Z})$, $\delta(\RV{T}, \RV{X}\RV{Y}.\RV{Y}\RV{Z})$, $\delta(\RV{T}, \RV{X}\RV{Y}.\RV{X}\RV{Z}.\RV{Y}\RV{Z})$, and $\Delta(\RV{T};\RV{Z}.\RV{X}\RV{Y})$ are invariant over $\Delta_P^{XY.XZ.YZ}.$
			\item $\delta(\RV{T},\RV{X}\RV{Z})$ since  $\CoI(\RV{T};\RV{X};\RV{Z})$, $\delta(\RV{T}, \RV{X}\RV{Y}.\RV{X}\RV{Z})$, $\delta(\RV{T}, \RV{X}\RV{Z}.\RV{Y}\RV{Z})$, $\delta(\RV{T}, \RV{X}\RV{Y}.\RV{X}\RV{Z}.\RV{Y}\RV{Z})$, and $\Delta(\RV{T};\RV{Y}.\RV{X}\RV{Z})$ are invariant over $\Delta_P^{XY.XZ.YZ}.$
			\item $\delta(\RV{T},\RV{Y}\RV{Z})$ since  $\CoI(\RV{T};\RV{Y};\RV{Z})$, $\delta(\RV{T}, \RV{X}\RV{Y}.\RV{Y}\RV{Z})$, $\delta(\RV{T}, \RV{X}\RV{Z}.\RV{Y}\RV{Z})$, $\delta(\RV{T}, \RV{X}\RV{Y}.\RV{X}\RV{Z}.\RV{Y}\RV{Z})$, and $\Delta(\RV{T};\RV{X}.\RV{Y}\RV{Z})$ are invariant over $\Delta_P^{XY.XZ.YZ}.$
		\end{enumerate}
		Hence the only partial information measure which is not fixed is $\delta(\RV{T};\RV{X}\RV{Y}\RV{Z})$ and
		\begin{equation}
			\begin{split}
		 		\Min_{\tiny Q\in\Delta_P^{XY.XZ.YZ}}\MI(\RV{T};\RV{X},\RV{Y},\RV{Z})	&= \delta(\RV{T}, \RV{X}\RV{Y}) + \delta(\RV{T}, \RV{X}\RV{Z}) + \delta(\RV{T}, \RV{Y}\RV{Z}) + \delta(\RV{T}, \RV{X}\RV{Y}.\RV{X}\RV{Z})\\
 			 											&+ \delta(\RV{T}, \RV{X}\RV{Y}.\RV{Y}\RV{Z}) + \delta(\RV{T}, \RV{X}\RV{Z}.\RV{Y}\RV{Z}) + \delta(\RV{T}, \RV{X}\RV{Y}.\RV{X}\RV{Z}.\RV{Y}\RV{Z})\\
 			 											&+ \delta(\RV{T}, \RV{X}) + \delta(\RV{T},\RV{Y}) + \delta(\RV{T},\RV{Z}) + \delta(\RV{T}, \RV{X}.\RV{Y}\RV{Z}) + \delta(\RV{T}, \RV{Y}.\RV{X}\RV{Z}) \\
 			 											&+ \delta(\RV{T}, \RV{Z}.\RV{X}\RV{Y}) + \delta(\RV{T}, \RV{X}.\RV{Y}) + \delta(\RV{T}, \RV{X}.\RV{Z}) + \delta(\RV{T}, \RV{Y}.\RV{Z})\\ 
 			 											&+ \delta(\RV{T}, \RV{X}.\RV{Y}.\RV{Z}).
			\end{split}
		\end{equation}
	The synergy is evaluate as
	\begin{equation*}
		\begin{split}
			\R{\CI}(\RV{T};(\RV{X},\RV{Y}),(\RV{X},\RV{Z}),(\RV{Y},\RV{Z}))	&= \MI(\RV{T};\RV{X},\RV{Y},\RV{Z}) - \Min_{\tiny Q\in\Delta_P^{XY.XZ.YZ}}\MI(\RV{T};\RV{X},\RV{Y},\RV{X})\\
										&= \delta(\RV{T}, \RV{X}\RV{Y}\RV{Z}).
		\end{split}
	\end{equation*}
	\subsection{Computing the Finest parts of the Trivariate PID}
	    The values of $\delta(\RV{T};\RV{X}), \delta(\RV{T};\RV{Y}), \delta(\RV{T};\RV{Z}), \delta(\RV{T};\RV{X}.\RV{Y}\RV{Z}), \delta(\RV{T};\RV{Y}.\RV{X}\RV{Z}),$ and $\delta(\RV{T};\RV{Z}.\RV{X}\RV{Y})$ can be extracted from the unique information $\R{\UI}(\RV{X}_k\backslash\RV{X}_i,\RV{X}_j)$
		of the subsystems of the form $(\RV{T},(\RV{X_i},\RV{X_j}),\RV{X_k})$ and $\R{\UI}(\RV{X}_k\backslash\RV{X}_i,\RV{X}_j)$ of the system $(\RV{T},\RV{X_i},\RV{X_j},\RV{X_k})$.
		
		 The synergy of the system $(\RV{T},\RV{X},\RV{Y},\RV{Z})$, the synergy of the system $(\RV{T},(\RV{X},\RV{Y}),(\RV{X},\RV{Z}),(\RV{Y},\RV{Z}))$, the synergy of the subsystems of the form $(\RV{T},(\RV{X_i},\RV{X_j}),\RV{X_k})$, and the synergy of the subsystems of the form  $(\RV{T},(\RV{X_i},\RV{X_j}),(\RV{X_k},\RV{X_\ell}))$ construct the following system of equations which allows to recover the individual synergistic quantities,
	\begin{equation*}
		\begin{split}
		\R{\CI}(\RV{T};\RV{X},\RV{Y},\RV{Z})	&=\delta(\RV{T};\RV{X}\RV{Y}\RV{Z}) + \delta(\RV{T};\RV{X}\RV{Y}) + \delta(\RV{T};\RV{X}\RV{Z}) + \delta(\RV{T};\RV{Y}\RV{Z})\\
												&+ \delta(\RV{T};\RV{X}\RV{Y}.\RV{X}\RV{Z}) + \delta(\RV{T};\RV{X}\RV{Y}.\RV{Y}\RV{Z}) + \delta(\RV{T};\RV{X}\RV{Z}.\RV{Y}\RV{Z})\\ 
												&+ \delta(\RV{T};\RV{X}\RV{Y}.\RV{X}\RV{Z}.\RV{Y}\RV{Z})\\
		\R{\CI}(\RV{T};(\RV{X},\RV{Y}),\RV{Z})	&= \delta(\RV{T};\RV{X}\RV{Y}\RV{Z}) + \delta(\RV{T};\RV{X}\RV{Z}) + \delta(\RV{T};\RV{Y}\RV{Z})\\
					 							&+ \delta(\RV{T};\RV{X}\RV{Z}.\RV{Y}\RV{Z})\\
		\R{\CI}(\RV{T};(\RV{X},\RV{Z}),\RV{Y})	&= \delta(\RV{T};\RV{X}\RV{Y}\RV{Z}) + \delta(\RV{T};\RV{X}\RV{Y}) + \delta(\RV{T};\RV{Y}\RV{Z})\\
					 							&+ \delta(\RV{T};\RV{X}\RV{Y}.\RV{Y}\RV{Z})\\
		\R{\CI}(\RV{T};(\RV{Y},\RV{Z}),\RV{X})	&= \delta(\RV{T};\RV{X}\RV{Y}\RV{Z}) + \delta(\RV{T};\RV{X}\RV{Y}) + \delta(\RV{T};\RV{X}\RV{Z})\\
					 							&+ \delta(\RV{T};\RV{X}\RV{Y}.\RV{X}\RV{Z})\\
		\R{\CI}(\RV{T};(\RV{X},\RV{Y}),(\RV{X},\RV{Z}))	&= \delta(\RV{T};\RV{X}\RV{Y}\RV{Z}) + \delta(\RV{T};\RV{Y}\RV{Z})\\
		\R{\CI}(\RV{T};(\RV{X},\RV{Y}),(\RV{Y},\RV{Z}))	&= \delta(\RV{T};\RV{X}\RV{Y}\RV{Z}) + \delta(\RV{T};\RV{X}\RV{Z})\\
		\R{\CI}(\RV{T};(\RV{X},\RV{Z}),(\RV{Y},\RV{Z}))	&= \delta(\RV{T};\RV{X}\RV{Y}\RV{Z}) + \delta(\RV{T};\RV{X}\RV{Y})\\
		\R{\CI}(\RV{T};(\RV{X},\RV{Y}),(\RV{X},\RV{Z}),(\RV{Y},\RV{Z}))	&= \delta(\RV{T};\RV{X}\RV{Y}\RV{Z}).\\
		\end{split}
	\end{equation*}
	
	Therefore, to compute the trivariate PID quantities then a hierarchy of a maximum entropy trivraite PID (Appendix~\ref{sec:apx-tri-pid}), six bivariate PID (Appendices~\ref{subsec:apx-sig-sys} and ~\ref{subsec:apx-dou-sys}), and a single optimization should be computed (Appendix~\ref{subsec:apx-trio-sys}). This hierarchy is scripted at the~\textsc{MaxEnt3D\_Pid}~\textsc{GitHub} in the file~\verb|test_trivariate_finer_parts.py|.
\end{appendix}
    
\end{document}